\documentclass[pre,superscriptaddress,a4paper,showpacs,twocolumn]{revtex4}
\usepackage{amsmath,graphicx,mathrsfs}
\usepackage{mathptmx,bm}
\pdfoutput=1

\begin{document}
\title{Monte Carlo study of multiply crosslinked semiflexible polymer networks}
\author{E.M. Huisman$^{1\ast}$, C. Storm$^{1,2}$, G.T. Barkema$^{1,3}$\\[3ex]
\normalsize{$^1$Universiteit Leiden, Instituut-Lorentz, Postbus 9506, NL-2300 RA Leiden, The Netherlands}\\
\normalsize{$^2$Department of Applied Physics and
Institute for Complex Molecular Systems, Eindhoven University of
Technology, P. O. Box 513, NL-5600 MB Eindhoven, The Netherlands}\\
\normalsize{$^3$Universiteit Utrecht, Institute Theoretical Physics, NL-3584 CE Utrecht, The Netherlands}
}

\begin{abstract}
We present a method to generate realistic, three-dimensional
networks of crosslinked semiflexible polymers. The
free energy of these networks is obtained from the force-extension
characteristics of the individual polymers and their persistent
directionality through the crosslinks. A Monte Carlo scheme is employed
to obtain isotropic, homogeneous networks that minimize the free energy, and for which all of the relevant parameters can be varied: the persistence length, the contour length as well as
the crosslinking length may be chosen at will. We also provide an
initial survey of the mechanical properties of our networks subjected to
shear strains, showing them to display the expected non-linear stiffening
behavior. Also, a key role for non-affinity and its
relation to order in the network is uncovered.\\
\\
$\ast$ electronic adress: lhuisman@lorentz.leidenuniv.nl

\end{abstract}


\maketitle

\section{Introduction}
Networks of semiflexible polymers have become one of the focal points in current soft matter research. The reason for this interest
is twofold: on the one hand, most relevant structural biological
materials, both intra- and extracellular, share the common architecture
of crosslinked semiflexible polymer networks. Two archetypical examples
are the cytoskeleton and the extracellular matrix.  At the same time,
there is a wide-spread realization that semiflexible networks represent
an interesting soft-matter system in their own right, outside of any
biological context, resulting in a much more fundamental interest in
the microscopic and geometrical origins of their mechanical behavior.

The mechano-elastic characteristics of networks of semiflexible polymers
have been studied to analyze and characterize different types of these
networks, both in vivo \cite{Sokolis}, and in vitro ~\cite{Kasza}. The
contributions of theory have been many and insightful, but analytical
progress has typically only been possible in certain limiting cases
where simplifying assumptions may be believed to hold, most notably
the assumption of affine deformations ~\cite{MacKintosh,Storm}. At the
same time, computer simulations have been used to study these networks,
but they too have had to rely on simplifications - either reducing
the system to two dimensions and limiting to the small-strain regime
\cite{Wilhelm2,Head} or ignoring the non-linear nature of the
constituent filaments \cite{Onck,Huisman}.

\begin{figure}[t] 
\begin{center}
\includegraphics[width=0.9\linewidth]{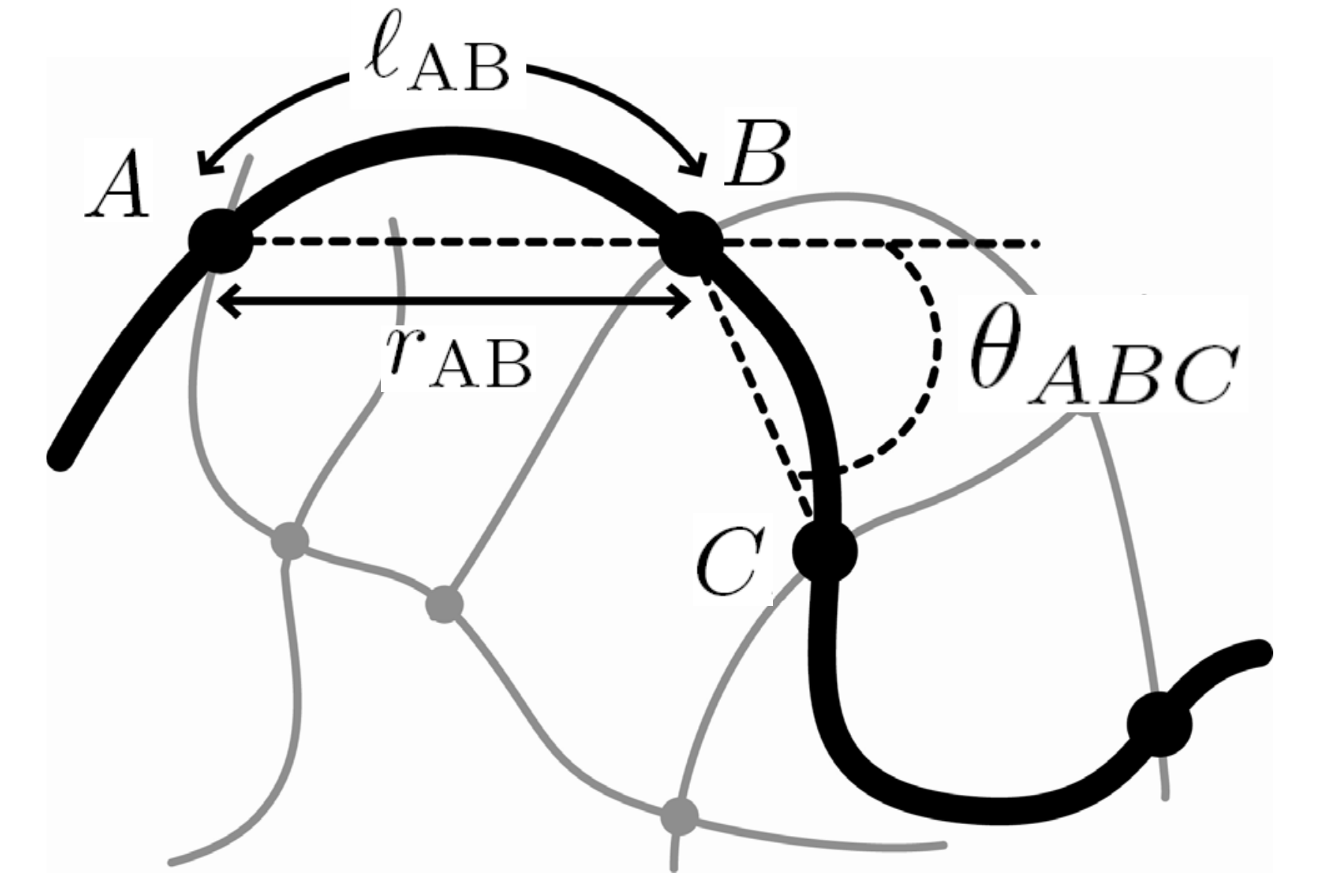}
\end{center}
\caption{{Schematic presentation of (part of) a semiflexible network, in which lines indicate the filaments and dots the crosslinks. The section of the filament between crosslinks $i$ and $j$ has length $l_{c,ij}$ and end-to-end distance $r_{ij}$. $\theta_{jkl}$ denotes the angle between two end-to-end vectors of neighboring segments along the same filament.}}
\label{fig1}
\end{figure}

We believe that the time is right for more realistic numerical modeling
of these networks that allows for a detailed microscopic look at the
relations between structure, geometry and mechanical properties. To this
end, we present a computer model to simulate these semiflexible polymer
networks in three dimensions. Networks are considered to consist of
filaments, described as semiflexible polymers.  These filaments are
crosslinked in various locations, which might induce extra bending of filaments, thus increasing the free energy of the system. We start with a homogeneous, isotropic
initial random network with a high free energy, and employ a Monte
Carlo scheme to relax this network. This approach allows us to generate
realistic three-dimensional networks containing hundreds of crosslinks,
which are nonetheless well equilibrated and thus represent realistic initial conditions for further mechanical loading in three dimensions.
The methodology to generate such networks is the first main result
presented in this paper, and is described in the first part of the paper (section II).

In the second part of this paper (section III), we subject these
networks to shear, and analyze their behavior as a function of the
network parameters, e.g. the stiffness and the length
of the filaments.  The results of these
computer experiments are compared with experiments to validate our model, and yield novel predictions for the mechanical behavior of
semiflexible networks.

\section{Network generation and equilibration}

We begin our discussion with a detailed look at the generation of our semiflexible networks based on single polymer energies, and how the Metropolis-Monte Carlo scheme is implemented and adapted for our specific purposes.

\subsection{Network Free Energy}

The networks considered in this manuscript consist of filaments,
which are linked by crosslinks $i=1\dots N_c$. Fig.~\ref{fig1} shows a schematic representation of a part of the network, indicating important parameters of the network and the notation used. Each
filament is an inextensible semiflexible chain, whose energy in the presence of an external force  is given by

\begin{equation}
E_{\rm fil}=\int_0^{l_c} \left(
                 \frac{\kappa}{2}
                    \left| \frac{d\hat{t}(s)}{ds} \right|^2
               + \frac{f}{2} \left| \hat{t}(s) \right|^2
               \right) ds,
\label{eq1}
\end{equation}
where $s$ is the arc length coordinate running along the filament, $\kappa$ is the
bending stiffness which is related to the persistence length $l_p$ as $\kappa=\beta^{-1} l_p$, with $\beta=1/(k_b T)$, $\hat{t}(s)$ is
the (unit) tangent vector along the filament, and $f$ is the applied force, directed along the end-to-end vector of the polymer.
The filamentous contribution to the total energy of a network is the sum of the energies of all filaments. In this work we consider inextensible filaments, thus ignoring backbone stretching of the filaments, a deformation that is only relevant at high forces for most biopolymers.

A brief note on our nomenclature: our networks consist of (multiply) connected {\em filaments}. Each of these filaments is partitioned into {\em segments}, which begin and end in {\em crosslinks}. A filament can thus consist of many segments, but is always a single mechanical entity, satisfying persistence not just at the segment level but also through crosslinks.

Each crosslink connects segments of two filaments, and our networks are therefore strictly tetrafunctional - albeit with the possibility of dangling ends which are discarded (we do not take steric avoidance into account).  Compared to the other scales in the network, crosslinks are assumed to be exceedingly small so that their only action, effectively, is to force a binary bond between two distinct filaments, or remote regions of the same filaments.

In our computer simulations, we store a complete list of all positions
$\vec{x}_i$ of the crosslinks, a complete list of the contour lengths $l_{c,ij}$ of the segments between crosslinks $i$ and $j$, and a connectivity table which lists which segments are linked by each of the crosslinks. We do not keep track of the spatial configuration of a segment between two crosslinks. Instead, we use the exact radial distribution function as computed from Eq. (\ref{eq1}) \cite{Wilhelm} to assign to each segment a contour length drawn from the radial distribution function computed at the segment's end-to-end length and persistence length. In this manner, we can already perform an important part of the full ensemble sampling in a straightforward manner: different assignments of the contour lengths correspond to different realizations of semiflexible networks with a prescribed spatial distribution of crosslinks. The relative likelihood of a given distribution of lengths is computed from the free energy of the resultant network, which we compute as follows.

\begin{figure} [t]
\includegraphics[width=0.95\linewidth]{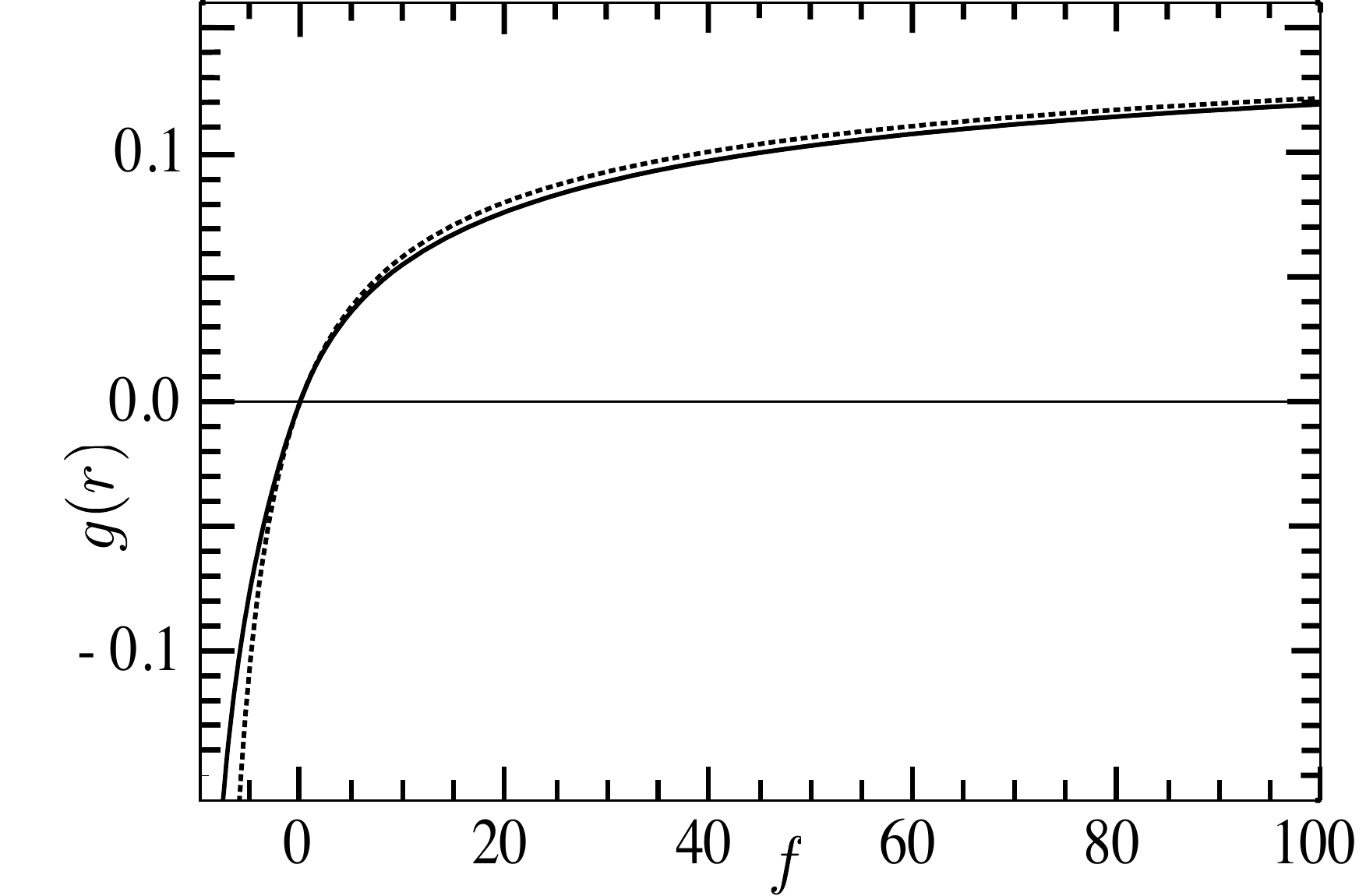}	
\includegraphics[width=0.9\linewidth]{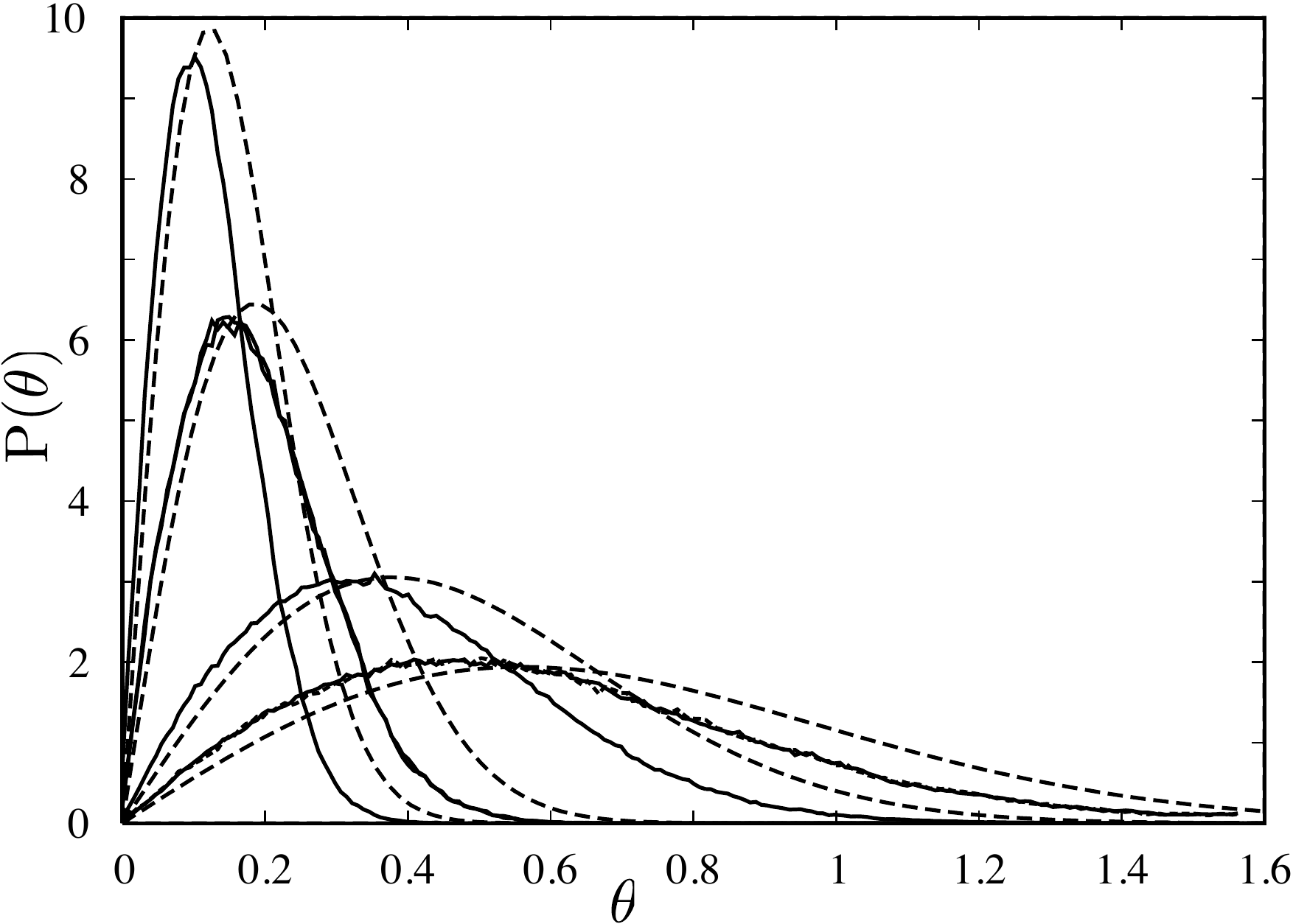}
\caption{(Color Online) {Validation of our effective Hamiltonian.} {a}) Analytic force-extension curve (straight line) (see \cite{Storm2}) vs. interpolation formula of force-extension (dotted line), where $f$ is the force on the segments and $g(r)$ the relative extension with respect to the equilibrium configuration. {b}) The probability distribution $P(\theta)$ of the angle $\theta$ between two segments, for different values of the length $l_1$, resp. $l_2$ of the first and second segment, measured in units of the persistence length $l_p$. In order of decreasing peak value, the solid lines correspond to $(l_1,l_2)=(l_p/100,l_p/50)$; $(l_p/100,l_p/16.7)$ coinciding with $(l_p/50,l_p/20)$; $(l_p/10,l_p/5)$; and $(l_p/10,l_p/16.7)$ coinciding with $(l_p/5,l_p/2)$.  The dotted lines are our theoretical approximations as given in Eq.~(\ref{eq2c}). Note that curves for constant $l_p/(l_1+l_2)$ fall on top of each other, both in the measured curves and in our approximation.}
\label{fig2}
\end{figure}

For a given network realization we partition the free energy in an internal segment part $F_2$ and an inter-segment part $E_3$. As stated above, the internal degrees of freedom of the segments are
integrated out. Thus, we express the free energy of a segment as a function of the distance between the crosslinks ($r_{ij}$) and the length
of the segment ($l_{c,ij}$). If the applied force $f$ in Eq. (\ref{eq1})
is positive (i.e., stretching the filament), $F_2$ can be computed from Eq. (\ref{eq1}) by employing a semiflexible analogue of the Marko-Siggia interpolation formula \cite{Marko}; an expression for this is given in the next section. The semiflexible WLC force-extension formula is not particularly accurate for negative forces, as the filaments quickly assume configurations with considerable transverse displacements under compressive loading. The crucial feature of compressive loading, however, is that the forces involved are always considerably smaller than those encountered for extensional loads - indeed, this asymmetry in the force-extension curve is responsible for many mechanical features of semiflexible networks. For negative forces, we find that the force-extension is adequately described by an exponential approach to the asymptote set by the classical Euler buckling force. Integrating the force-extension curve yields the following expression for the energy

\begin{equation}
\beta F_2=\left\{\begin{array}{cl} -\frac{9g(r_{ij})^2(5+6g(r_{ij}))}{-1+6g(r_{ij})} \qquad \qquad & \text{if }\; f>0\\
\\
                 |(-\frac{1}{90}(-1+\exp(90g(r_{ij})/\pi^2))\pi^4+\pi^2)| & \text{if }\; f<0\\
                 \end{array} \right.
\label{eq2a}
\end{equation}
where $g(r_{ij})$ is the scaled extension given by
\begin{equation}
g(r_{ij})=-l_p/l_{c,_{ij}}+1/6+l_pr_{ij}/l_{c,_{ij}}^2.
\end{equation}
These equations are not only computationally convenient, they also provide an excellent fit to the full, analytical force-extension curves as shown in Fig.~\ref{fig2}a, where we plot the force vs. the scaled extension $g(r_{ij})$. In addition to the single-segment force-extension, we also need to keep track of their persistence through crosslinks. There is no analytical formula for this contribution, and we have therefore simulated many individual filaments to obtain a reliable numerical expression for this contribution. If the applied force $f$ in Eq. (\ref{eq1})
is positive (i.e., stretching the filament), it turns out that we can capture the essential behavior by

\begin{equation}
\beta E_3 =\frac{l_p\theta_{ijk}^2}{l_{c,{ij}}+l_{c,{jk}}},
\label{eq2c}
\end{equation}
where $l_{c,_{ij}}$ and $l_{c,_{jk}}$ are the contour lengths of the segments
and $\theta_{ijk}$ is the angle between the two end-to-end vectors of the
segments. Note that this contribution to the total energy is not accompanied by an entropic contribution, since it is defined by explicit variables in our network.

To assess the quality of the segment-segment energy function, we compare the distribution function of this energy with simulations. We simulate a single wormlike
chain of length $L_w$, at a fixed temperature and a persistence
length $l_p$, and count the probability $P_{\rm wlc}(\theta)$ of an
angle $\theta$ between the vectors $\vec{r}_{L_w} - \vec{r}_{N}$
and $\vec{r}_{N} - \vec{r}_1$. Here, $N$ is anywhere on the chain. Our approximate expression for this
probability is $P_{\rm app}(\theta) \sim N(lp/(l_{c_1}+l_{c_2})) \theta \exp(-\beta E_3(\theta))$, in which
the energy $E_3$ is given by Eq. (\ref{eq2c}) and $N(l_p/(l_{c_1}+l_{c_2}))$ is a normalization factor.
This histogram is plotted in Fig.~\ref{fig2}b. Although the correspondence is not perfect, this formula does reflect the essentials of the angle distribution, capturing the broadening and shift of its peaks.

In summary, we attribute to a specific network configuration an
energy which is the sum of single-segment energies given by
Eqs. (\ref{eq2a}), plus a sum over all segment-pair
energies given by Eq. (\ref{eq2c}), which runs over all pairs of segments
belonging to the same filament and meeting in the same crosslinks.

\subsection{Interpolation Formula for the Segment Free Energy}
Eq. (\ref{eq1}) enables us to derive an analytic approximation for the
semiflexible force-extension relation. To simplify notation,
we will pass to dimensionless quantities, rescaling all forces by a factor of $l_c^2/\kappa$ and all lengths by $l_p/l_c^2$. Based upon Eq. (\ref{eq1}), we can express the
scaled difference between the total rescaled length of the polymer ($\tilde{l}_c$)
and the end-to-end length at rescaled force $\phi$ ($\tilde{l}_{\phi}$) as \cite{Storm}:

\begin{equation}
\tilde{l}_c-\tilde{l}_{\phi}=\frac{1}{\pi ^2}\sum_{n=1}^{\infty}\frac{1}{n^2+\phi},
\label{eq2}
\end{equation}
which gives:
\begin{equation}
\tilde{l}_c-\tilde{l}_{\phi}=\frac{-1+\sqrt{\phi}\coth\sqrt{\phi}}{2\phi}.
\label{eq3}
\end{equation}
At zero force this gives $\tilde{l}_c-\tilde{l}_0=1/6$. With this, we
can define the differential extension at force $\phi$ (i.e, the incremental extension compared to that at zero force) as $\delta \tilde{l}=\tilde{l}_\phi-\tilde{l}_0$. We use these equations to construct an interpolation formula for $\phi(r,l_c,l_p)$, which is the direct analogue of the Marko-Siggia interpolation for the WLC\cite{Marko}. Around $\tilde{l}_0$ Eq. (\ref{eq3}) gives as a first order approximation:
\begin{equation}
\phi=90\delta \tilde{l}.
\label{eq4a}
\end{equation}
In the large force regime we can expand Eq. (\ref{eq2}) to yield
\begin{equation}
\phi=\frac{1}{4(1/6-\delta \tilde{l})^2}.
\end{equation}
Tying the two asymptotes together yields
\begin{equation}
\phi=-18\delta \tilde{l} + \frac{1}{4(1/6-\delta \tilde{l})^2} -9,
\label{eq4}
\end{equation}
which can be integrated once to yield Eq. (\ref{eq2a}). Fig.~\ref{fig2}a shows the comparison between this formula and the exact solution - the difference between the two does not exceed $6\%$.

\subsection{Network generation}
\begin{figure}[t] 
\begin{center}
\includegraphics[width=0.9\linewidth]{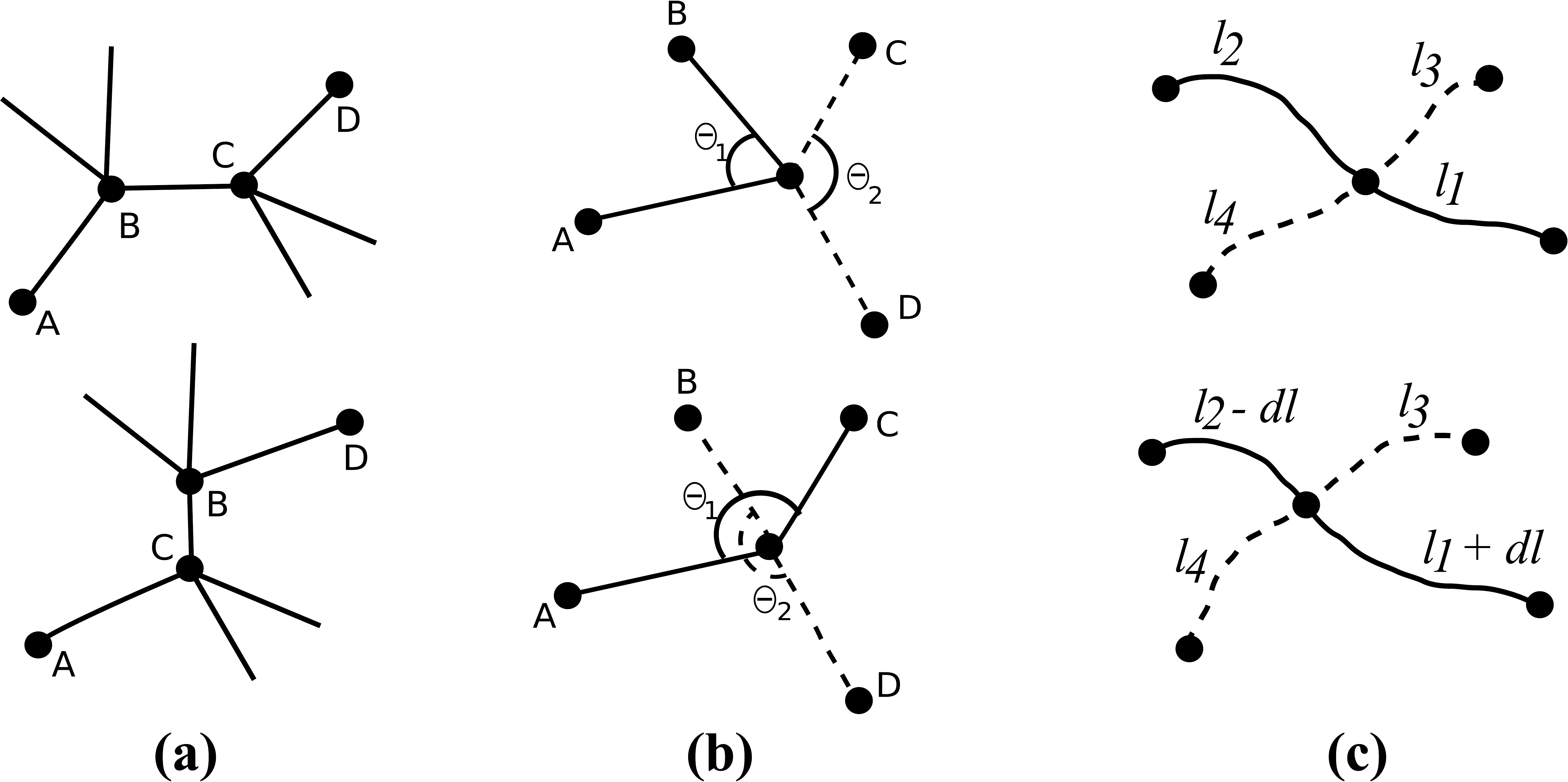}
\end{center}
\caption{{Schematic representation of the three Monte Carlo moves.} {a)} Four crosslinks
that are connected as shown in the above figure, are randomly
selected in the network. Bonds AB and CD are broken and bonds AC and BD
are created, such that the configuration of the lower figure is formed
after energy relaxation. {b)} A crosslink is
randomly chosen at which crosslinks A and B are part of the same filament,
as are crosslinks C and D (above figure). Now A and C become part of
the same filament as do B and D. This alters the three-crosslink free energy, $E_3$. {c)} A randomly chosen length ($dl$) is removed from the length of one
segment of a filament and transferred to a neighboring segment of the
same filament, such that the configuration of the lower figure is formed
after relaxation.}
\label{fig3}
\end{figure}

\begin{figure}[t] 
\begin{center}
\includegraphics[width=0.8\linewidth]{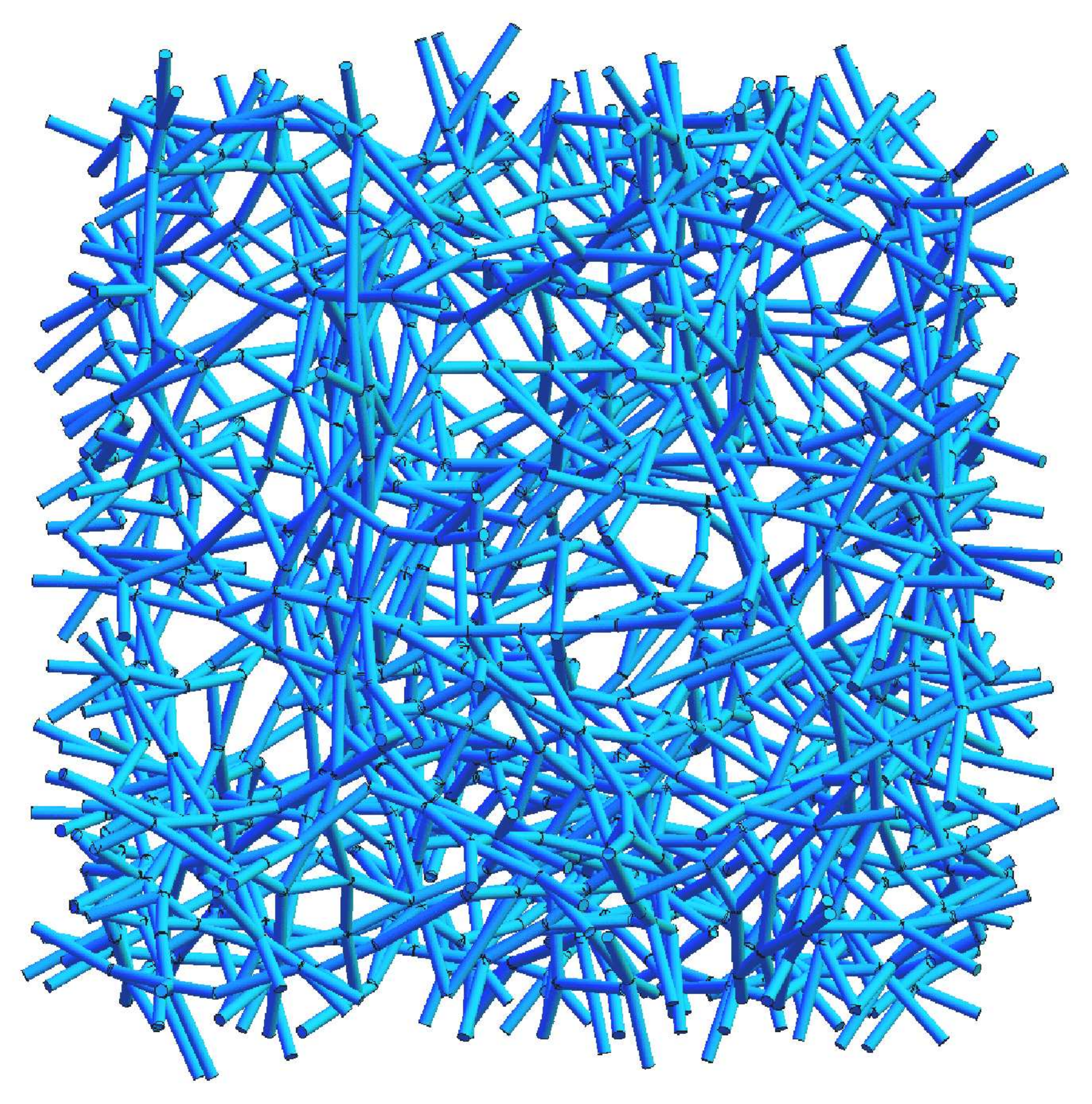}
\end{center}
\caption{{Representation of a generated network}. This network consists of 333 filaments, each on average crosslinked six times. The network is periodic in all three directions. Note that the undulations of the segments are not represented.}
\label{fig4}
\end{figure}

The task at hand is obviously to determine network configurations that minimize the free energy thus defined. To this end, we use a Monte Carlo minimization scheme: starting from an isotropic, random network, we propose random changes in topology, each of which is either accepted or rejected according to the Metropolis criterion.

The initial network is constructed by
placing $m$ nodes with random coordinates in a cubic periodic cell. To
connect these nodes into a four-fold coordinated network, we proceed iteratively: we begin by identifying three nodes which are
close to each other and connect these with a loop of three bonds.
This loop is then extended one bond at a time: we identify a node $A$
which is not fully connected and which is closest to an existing bond $BC$,
and then replace this existing bond by two bonds $AB$ and $AC$. This process
is repeated until all nodes are four-fold connected.

In the resulting fully fourfold-coordinated network, each bond is
considered to be a segment of a single, long filament. This network as a whole can therefore be considered a single, circular
filament which is crosslinked to itself at various places. We then proceed to minimize the free energy - computed as detailed before - of
this initial network, using the standard local minimization method of damped
molecular dynamics.

The initial network will be highly stressed, and in general far removed from a realistic equilibrium configuration. Chiefly, this is due to considerable filament bending, with intra-filament bends at crosslinks often exceeding 90 degrees.  As initial large strides towards an optimal configuration will proceed along downhill directions related to the release of precisely these dominant bending stresses, we first focus on rearranging the topology of the network, analogous to the continuous random network approach, pioneered by Wooten, Winer, Weaire \cite{www} and further extended and optimized as detailed in \cite{barkema}. This is realized by a series of Monte Carlo moves that alter the topology; these are moves (a) and (b) in Fig.~\ref{fig3}. To the initial configuration with a
topology $L'$ with minimized coordinates $\vec{x}'$, we assign a free
energy $F'$ as obtained from Eq. (\ref{eq2c}) plus a quadratic function around the
average bond distance, to prevent crosslinks from clustering and to tune the final network topology. The average bond distance determines whether the final network will be densely or loosely crosslinked. We then change the topology to $L''$ by one of the moves, and relax
the network with this new topology, resulting in the new coordinates
$\vec{x}''$ and a new free energy $F''$. Depending on the change in
free energy $\Delta F = F''-F'$, the topological change is accepted
or rejected, using the Metropolis algorithm. Note that in this stage,
we assume that the free energy of a network with minimized crosslinks
coordinates is representative for the free energy of all networks with the
same topology, up to some additive constant that is topology-independent.

Once such topology altering moves no longer significantly affect the overall energy - this typically happens in configurations where the bending angle of
the filament in each node is on average around 20 degrees - contour lengths are
attributed to the segments. As explained, for a segment $AB$ with end-to-end distance $r_{AB}$, the length $l_{c,AB}$ is drawn from the corresponding distribution
for the WLC with the desired persistence length $l_p$.  Next, we chop up the single continuous filament into many smaller ones, by random deletion of segments under the constraint that all crosslinks stay connected,
up to the point where the desired number of filaments (or, alternatively, mean filament length) is reached. This network is then further equilibrated
with the Monte Carlo moves (b) and (c) shown in Fig.~\ref{fig3}, each of which is now accepted to a comparable degree. To avoid computational instabilities for floppy filaments we add a short-range repulsive force between crosslinks. A typical network generated with this approach is shown in Fig.~\ref{fig4}.

\section{Mechanical Response of the Network}

The ultimate goal is to understand the relationship between the structure of a network and its mechanical properties. In the following sections, we explore some of the basic mechanical properties of our system in an attempt to check whether well-known behavior is correctly reproduced, and simultaneously to offer a glimpse of the relevant microscopic processes that we are now able to study in detail and their role in the overall mechanics.

\begin{figure}[t] 
\includegraphics[width=0.7\linewidth]{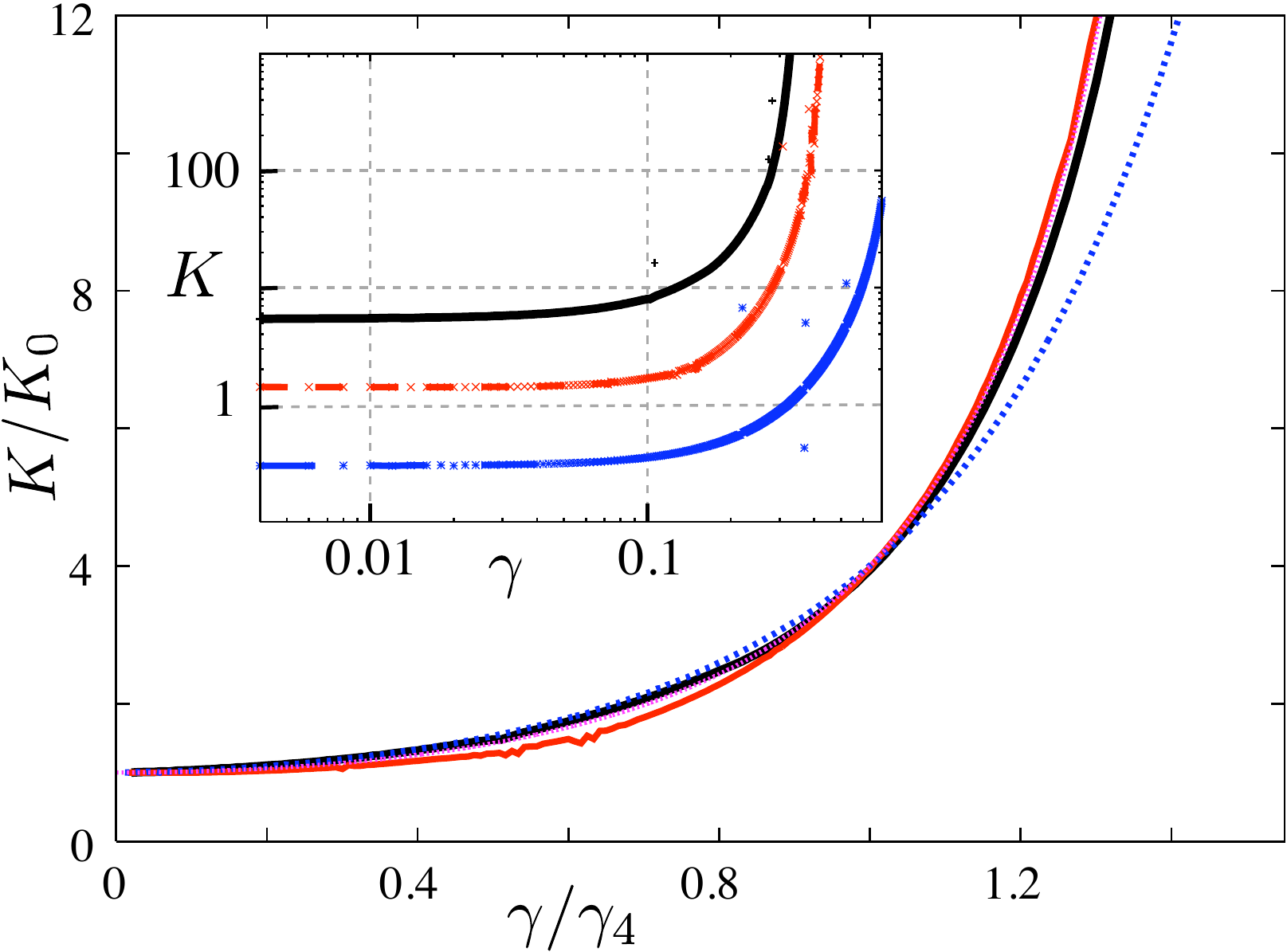}
\includegraphics[width=0.7\linewidth]{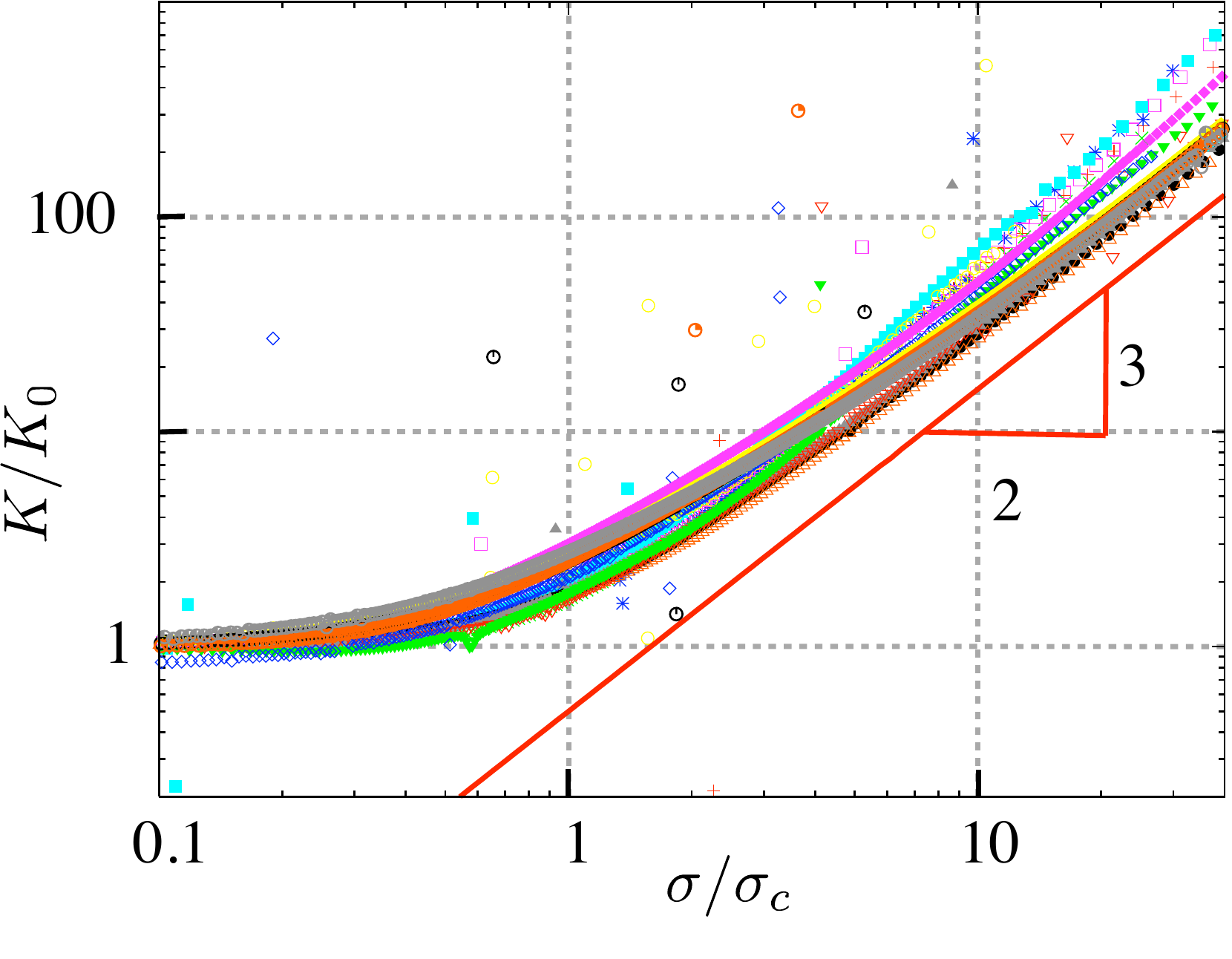}
\caption{(color online) {a)} Master curve of scaled differential modulus as a function of scaled shear-strain $\gamma/\gamma_4$, where $\gamma_4$ is the strain at which the modulus is four times the initial modulus $G_0$. For the curves shown, the average number of crosslinks per filament is $L/l_c=6$. The values for the scaled persistence length $l_p/l_c$ used are $15.7$, $3.81$ and $0.77$, of which only the latter is distinguishable at large strains (dotted blue line). Besides, we plotted the scaling from affine theory, which overlaps with the other curves. The inset shows the original strain-stiffness curves, from top to bottom with scaled persistence lengths of $15.7$, $3.81$ and $0.77$. Note that we plot all data points and draw a curve through them. However, a couple of the data points lie well outside the curve (see sect. 2.3). {b)} Differential modulus of the networks as a function of the scaled shear stress $\sigma$ for 18 networks with varying $l_p/l_c$ and $L/l_c$.}
\label{fig5}
\end{figure}

\begin{figure}[t] 
\includegraphics[width=0.7\linewidth]{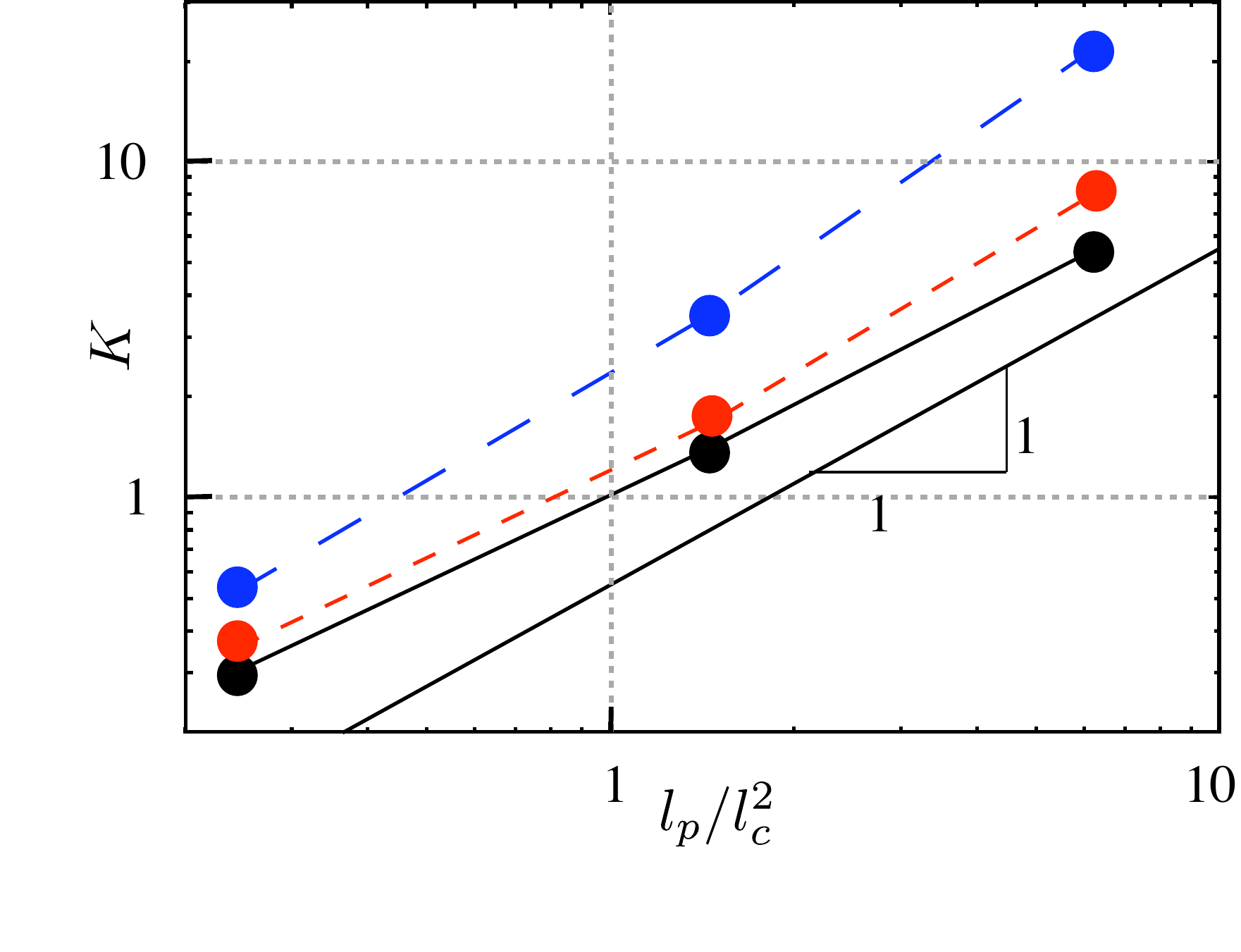}
\includegraphics[width=0.7\linewidth]{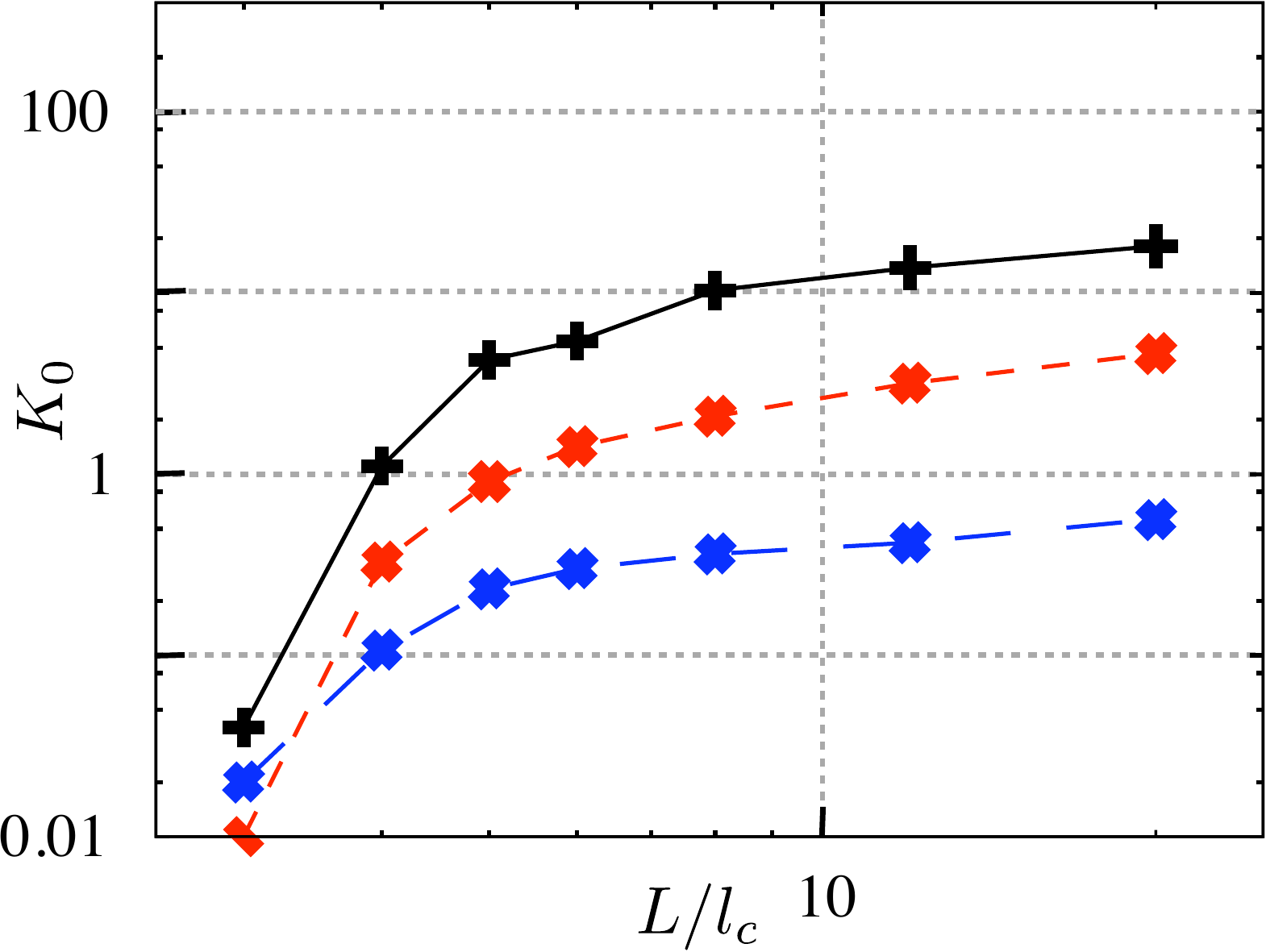}
\caption{(color online) {a)} Differential modulus $K$ as a function of $l_p/l_c^2$. The three curves correspond to different strains $\gamma$, from bottom to top equal to $0.02$, $0.10$ and $0.20$. For all curves shown, $L/l_c=6$. {b)} $K$ at $\gamma=0$ as a function of $L/l_c$ for various $l_p/l_c$ (upper line: $l_p/l_c=15.7$, middle line: $l_p/l_c=3.81$, bottom line: $l_p/l_c=0.77$).}
\label{fig6}
\end{figure}

The behavior of biopolymer networks under strain depends on many experimental parameters, such as the concentration of biopolymer, the amount of capping proteins and the concentration and characteristics of binding proteins. In our simulations, we can reproduce such changes by varying the crosslinking length, the persistence length and the average number of crosslinks per filament. In this paper we consider networks that consist of $10^3$
crosslinks connecting $2.10^3$ segments. Periodic boundary conditions are applied in all three directions. We do not take into account the
contributions of the dangling ends of filaments, nor do we consider the
excluded volume. Our networks are typically very densely crosslinked,
which implies that filaments which are close to each other have a high
probability to be crosslinked, and we therefore feel it is justified to
assume the crosslinking constraints to dominate over the effect of
entanglements. Additionally, most naturally occuring biopolymer
networks, as well as most {\em in vitro} biomimetic
experiments considered occur at fairly low polymeric volume fractions, further
reducing the importance of excluded volume. The persistence length in the networks we use ranges from $l_p/l_c=1$ to $l_p/l_c=16$ and the average length of filaments from $L/l_c=3$ to $L/l_c=20$. The data graphed in this paper is obtained from single, representative
networks as large as practically possible to minimize the effects of
finite system size on the observables that we measure. 

The experimental techniques used to probe the mechanical response are essentially twofold: on the one hand, in vitro networks are often subjected to global shears in commercial rheometric setups to characterize their macroscopic visco-elastic properties \cite{Janmey,Liu}. On the other hand, many experiments focus rather on the microscopic processes involved by injecting small particles ($\sim 1 \mu m$) in the network to monitor the behavior at the filament scale of the network \cite{Koenderink,Janmey}. Our computational method allows us to work at both levels by direct and simultaneous measurement of the overall stiffness as well as all individual displacements and forces in the system, to high accuracy.

We model shearing by virtually displacing all crosslink positions affinely by small shear-increments of 0.2\%. After each shear-increment we allow for non-affine relaxation of all individual crosslinks in order to minimize the free energy of the network. During this procedure the forces and displacements are recorded and can be used for further analysis of the network response. It's important to note that we allow for full relaxation after each strain increment - this would be appropriate for adiabatically slow shears and should therefore be compared to the zero-frequency limit in oscillatory rheology, as indeed we shall do.

\subsection{Strain Stiffening}

To characterize our networks, we first consider the differential stiffness, $K=\partial \sigma / \partial \gamma$ during shear. An important and characteristic feature of these networks is their highly non-linear stiffening behavior under relatively small shear stresses \cite{Janmey}. As argued in Ref. \cite{Storm}, all experimental curves of the modulus of networks of semiflexible polymers collapse for small shears on a master curve by scaling the stiffness by the initial stiffness ($K/K_0$) and scaling the shear by its value at which the stiffness is four times the initial stiffness ($\gamma / \gamma_4$). Fig.~\ref{fig5}a shows the scaled strain-stiffness curves of our networks under shear, where we plotted the differential modulus $K$, as a function of shear for different ratio's between $l_p$ and $l_c$ (the average contour length of the segments). We observe the same universal scaled stiffening behavior as observed in experiments. For comparison, we plotted the theoretical curve that incorporates the typical force-extension curve of single filaments combined with the assumption that the filaments deform affinely, that has shown to represent this same master curve. Note that we do not account for rupture and backbone stretching of the filaments, which becomes relevant at larger shears. The inset shows the original curves, where one clearly sees an increase in the initial stiffness as well as a small decrease in the strain at which the networks start to stiffen by increasing the stiffness of individual filaments. In our simulations, $l_p/l_c \approx 16$ is more or less comparable with an actin network with an average distance between crosslinks of 1 $\mu m$ and an average filament length of 6 $\mu m$. Smaller values of $l_p/l_c$ represent networks of filaments with a lower persistence length like fibrin or networks that are less dense.

Another way to compare our results with experiments is to look at the scaling in the large strain limit. By superposition of a small oscillatory stress on a prestress, the differential modulus can be experimentally measured. From these measurements it is known that $K\sim \sigma^{1.5}$ for large stresses $\sigma$ \cite{Gardel}. We plotted $K/K_0$ vs. $\sigma/\sigma_{\rm c}$, where $\sigma_{\rm c}$ is the critical stress, defined as the intersection between the horizontal low-stress regime and the high-stress asymptote. As shown in Fig.~\ref{fig5}b, all our networks show the same characteristic scaling behavior at large shears. Combined with the observed stiffening, this indicates that we capture the essential physics in our model, both at small and large shears.

Eq. (\ref{eq4a}) indicates that the initial stiffness of individual filaments scales as $K_{0,\rm fil}\sim l_p^2/l_c^4$. We expect this scaling behavior to change when the filaments are placed in a network that allows non-affine reorientations, as is the case in our simulations. Since all filament properties scale with $l_p/l_c^2$, we plot $K$ vs $l_p/l_c^2$, see Fig.~\ref{fig6}a. The figure shows that $K_0\sim (l_p/l_c^2)^1$ which emphasizes that for non-affine deformations the persistence length of the constituent filaments is less important for the overall network behavior, as filament reorientations allow for an alternative route to comply with the imposed strains. For increasing $\gamma$, the steepness of the slope increases, which strongly correlates with the stiffening of the networks.

From experiments \cite{Koenderink,Jliu} and simulations \cite{Huisman,Heussinger} it is known that the average filament length influences the network response. In cells many capping proteins are active that can control the length of the filaments, thus changing the mechanical properties. We measured the initial stiffness $K_0$ as a function of the average filament length $L$ and $l_p$. $L/l_c$ can be considered as the average number of crosslinks on a filament, which we can vary while keeping $l_c$ constant. As expected, Fig.~\ref{fig6}b shows a decrease in $K_0$ if the average filament length decreases. Segments of the same filament influence each others displacements, thus restricting the freedom to adapt to stresses. Besides, when crosslinks connect two or three segments instead of four, these crosslinks are more flexible to reorient when sheared. Therefore, networks with short filaments are softer during shearing.

Fig.~\ref{fig6}b also shows that the stiffness becomes nearly zero for short filaments, a behavior independent of $l_p$. This decrease is related to the percolation of the network. When the filaments become too short, no real network will be formed. In that case, shearing will shear the liquid in which the filaments are immersed, but the filaments will not be constrained in their movement and thus the stiffness will vanish. Please note that we employ a specific procedure to remove material from the network to generate increasingly sparse networks, which implies that the filament length at which the modulus vanishes cannot be directly related to experiments. The overall trend, however, is representative of real networks.

\subsection{Non-Affine Behavior and Ordering}

While the system deforms in our simulations, we allow for non-affine reorientations of the segments. It has been suggested that such non-affine deformations greatly alter the mechanical response \cite{Onck}, and indeed we find that this is true. First, a few words about the definition and measures of non-affinity. In general, an applied macroscopic strain maps any material point $\bm x$ in the reference space in the network onto a new point $\bm x'$ in the target space. The location of the point in the target space may be thought of as arising from a combination of an affine deformation and a non-affine contribution:

\begin{equation}
{\bm x} '=\Lambda(\gamma){\bm x}+{\bm \Delta}({\bm x},\gamma)\quad ,
\end{equation}

where $\Lambda(\gamma)$ is the deformation gradient tensor, which for the case we consider - three-dimensional simple shear in the $\hat x_1$-direction - is given by
\begin{equation}
\Lambda(\gamma)=\left( \begin{array}{ccc} 1&\gamma&0\\0&1&0\\0&0&1\end{array} \right).
\end{equation}
The observation that the non-affine contribution ${\bm \Delta}({\bm x},\gamma)$ depends both on the applied strain and the (original) location of the point under consideration immediately raises the question of what, precisely, it means for a system to be affine. In the strictest sense, an affine system may be defined as one obeying ${\bm \Delta}({\bm x},\gamma)=0$ for all ${\bm x},\gamma$. This definition, however, is highly restrictive as it does not allow for {\em any} non-affine motion at any point. For systems that do behave non-affinely to some extent, the most general measure of the extent of this non-affinity was shown in \cite{didonna} to be the non-affinity correlation function
\begin{equation}
{\sf A}_{ij}({\bm x},{\bm x'};\gamma,\gamma')=\langle \Delta_i({\bm x},\gamma) \Delta_j({\bm x'},\gamma')\rangle,
\end{equation}
where the average $\langle \dots \rangle$ is over all crosslinking points in the network. Both its spatial dependence and the strain dependence are of interest - in a moment we will investigate the strain dependent aspects by focusing on the trajectory that single points trace out during a deformation. Following \cite{didonna}, we shall measure to this end the equal argument limit of the trace of ${\sf A}_{ij}({\bm x},{\bm x'};\gamma,\gamma')$ which we shall call simply ${\sf A}(\gamma)$:
\begin{equation}
 {\sf A}(\gamma)=\frac{1}{\gamma^2}\langle |{\bm \Delta}({\bm x},\gamma)^2|\rangle.
\end{equation}
Note that this measure need not approach zero at large strains, even though one may expect all segments to become aligned with the direction of maximal strain in this limit and experience, in effect, a purely extensional strain. The reason ${\sf A}(\gamma)$ does not tend to zero lies in the fact that even though the deformation becomes {\em differentially} affine, it does not become affine in the absolute sense. To focus on this differential affinity, which we feel is a more appropriate measure of (asymptotic) affinity, we introduce a second measure by considering the differential displacement from the initial point ${\bm x}_i$ to the final point ${\bm x}_f$ before and after a small strain increment ${\rm d}\gamma$:
\begin{equation}
{\bm x}_f=\Lambda({\rm d} \gamma){\bm x}_i+\tilde{\bm \Delta}({\bm x}_i,\gamma+{\rm d} \gamma).
\end{equation}
We use this to define the differential non-affinity measure
\begin{equation}
\delta {\sf A}(\gamma)=\frac{1}{({\rm d}\gamma)^2}\langle |\tilde{\bm \Delta}({\bm x},\gamma)^2|\rangle.
\end{equation}
This measure {\em does} go to zero as $\gamma$ becomes very large. Later, our simulations will show that we do not expect this limit to be attained in experiments as, for realistic parameter values, the system will have failed long before. ${\sf A}$ and $\delta {\sf A}$ may be expressed in terms of each other, and the latter tending to zero implies that asymptotically, ${\sf A}$ should become constant with the magnitude of this constant reflecting the overall strength of the past non-affinity.

\begin{figure}[t] 
\begin{center}
\includegraphics[width=0.7\linewidth]{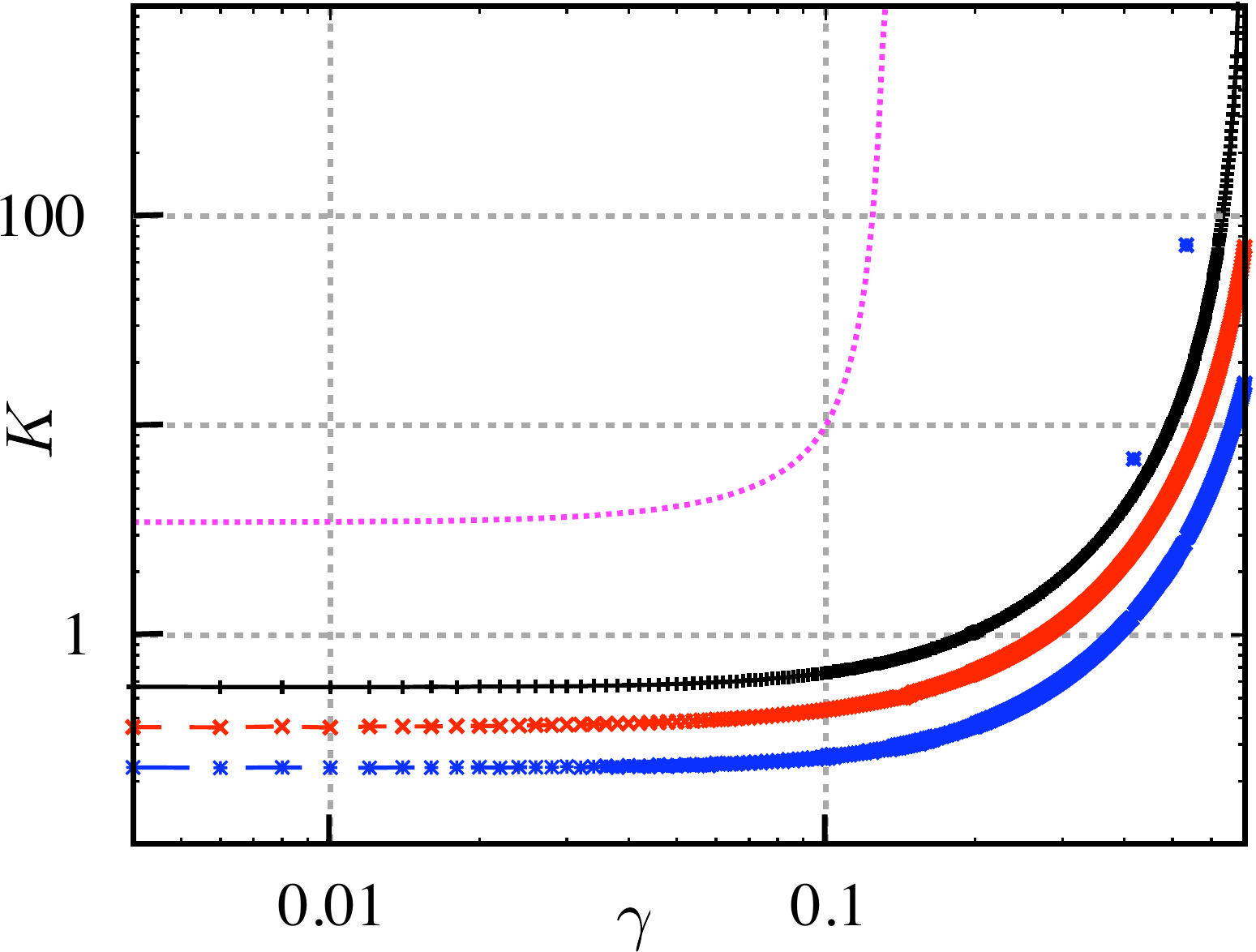}
\includegraphics[width=0.7\linewidth]{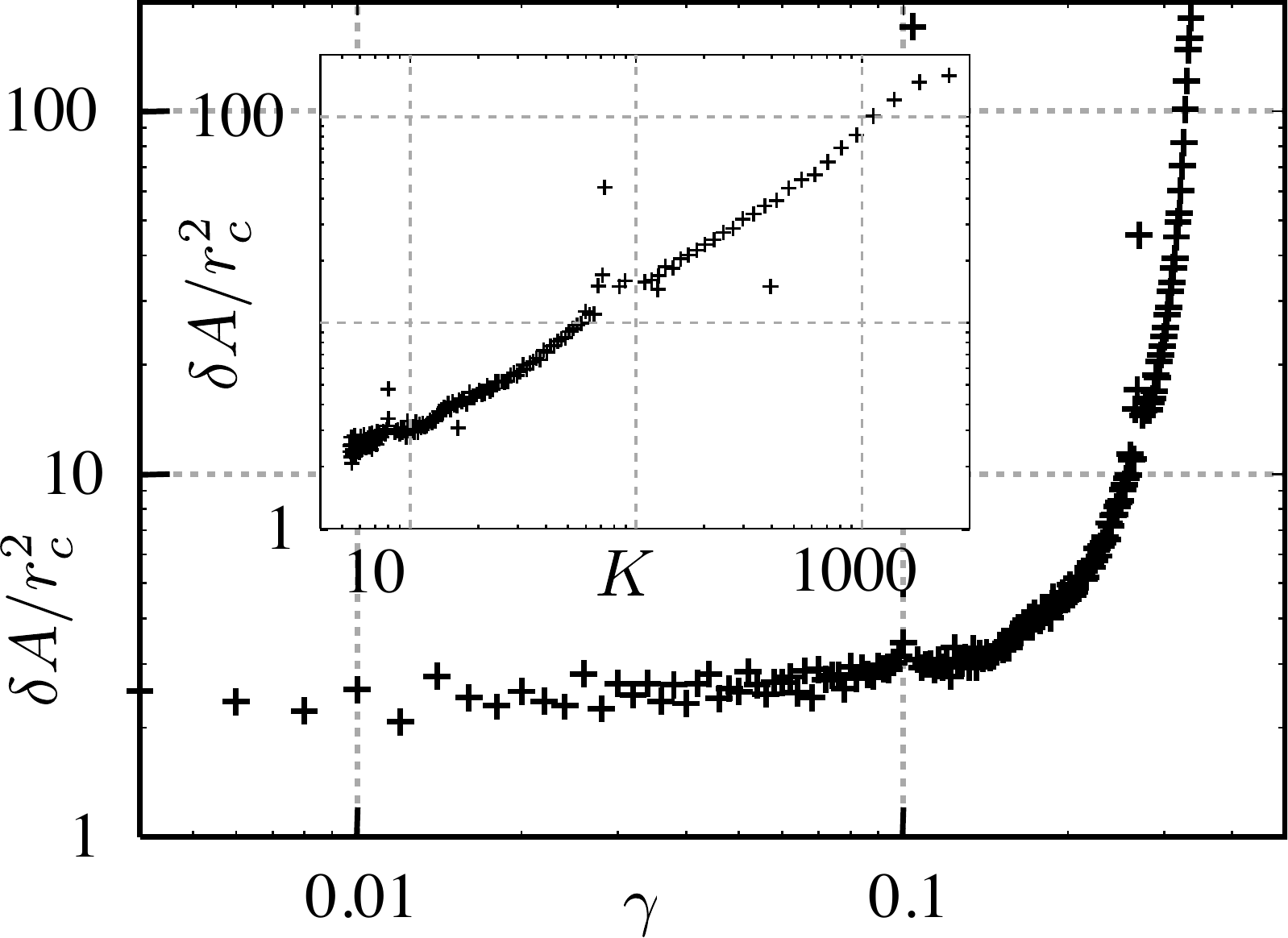}
\end{center}
\caption{(color online) Non-affine behavior of the networks during shearing. {a)} The differential modulus $K$ measured during affine deformation (upper, dotted line) and during non-affine deformation (from bottom to top: $L/l_c=5.0$, $L/l_c=8.0$ and $L/l_c=20.0$). For all networks, $l_p/l_c=0.77$. {b)} Differential non-affinity $\delta {\sf A}$ as a function of strain for different $l_p/l_c$. For all curves, $L/l_c=6$. To relate $\delta {\sf A}$ to other length scales in the system, we plot $\delta {\sf A}/r_c^2$, where $r_c$ is the average distance between crosslinks. A value of $1.0$ implies that the average non-affine displacement is equal to $r_c$ if $\gamma$ would be $1.0$. The inset shows the stiffness vs. the differential non-affinity.}
\label{fig7a}
\end{figure}

Ultimately, we are interested to see to what extent non-affinity affects the mechanical response. To monitor this influence, we perform a shear without relaxation after each strain increment, thus obtaining $K_\mathrm{affine}$. Fig.~\ref{fig7a}a shows both $K_{\rm affine}$, which is independent of $L/l_c$, and $K$ for networks with different filament length, all having $l_p/l_c=1$. As can be seen, even for long filaments, the difference between affine deformation and non-affine deformation is striking, both for he initial modulus $K_0$ and for the onset of stiffening. This puts the so-called linear ({\em i.e.}, small-strain) regime of network elasticity in a new perspective: even though the strains are small, there is always a finite amount of non-affinity which greatly affects the overall small strain response. It is thus crucial to understand the role of non-affinity, even at small strains, to predict the network modulus.

Even though the Hamiltonian of the network remains the same, the difference between the strain-stiffness curves is striking. These differences can only be due to non-affine behavior of the network, as all other determinants - topology, filament length, density and persistence length, are identical. There has been some debate whether the origin of stiffening is ultimately entropic or mechanical, but our results suggest that rather, we should focus our attention on the degree of non-affinity which acts to delay and attenuate the stiffening.

\begin{figure}[t] 
\begin{center}
\includegraphics[width=0.7\linewidth]{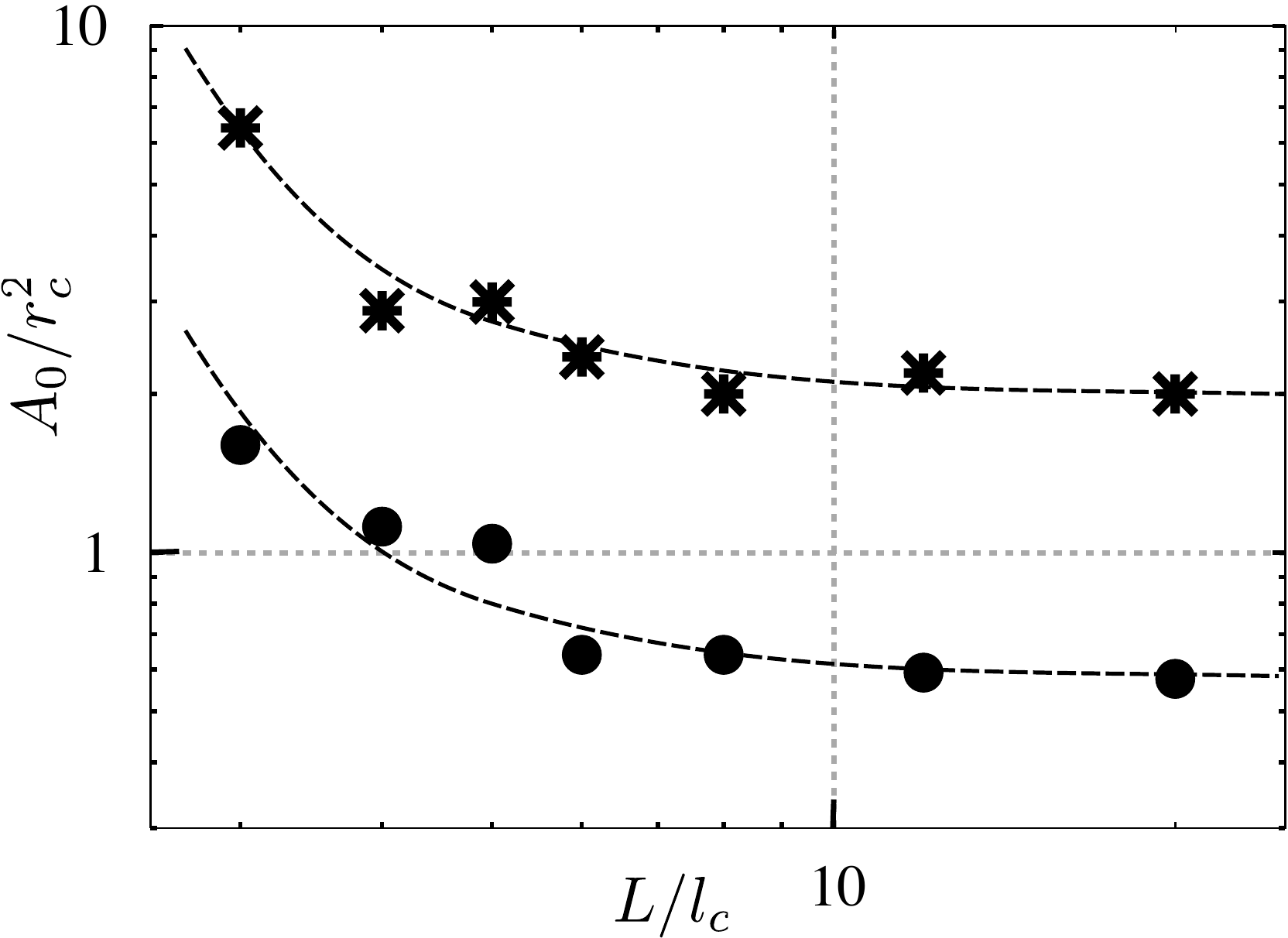}
\includegraphics[width=0.7\linewidth]{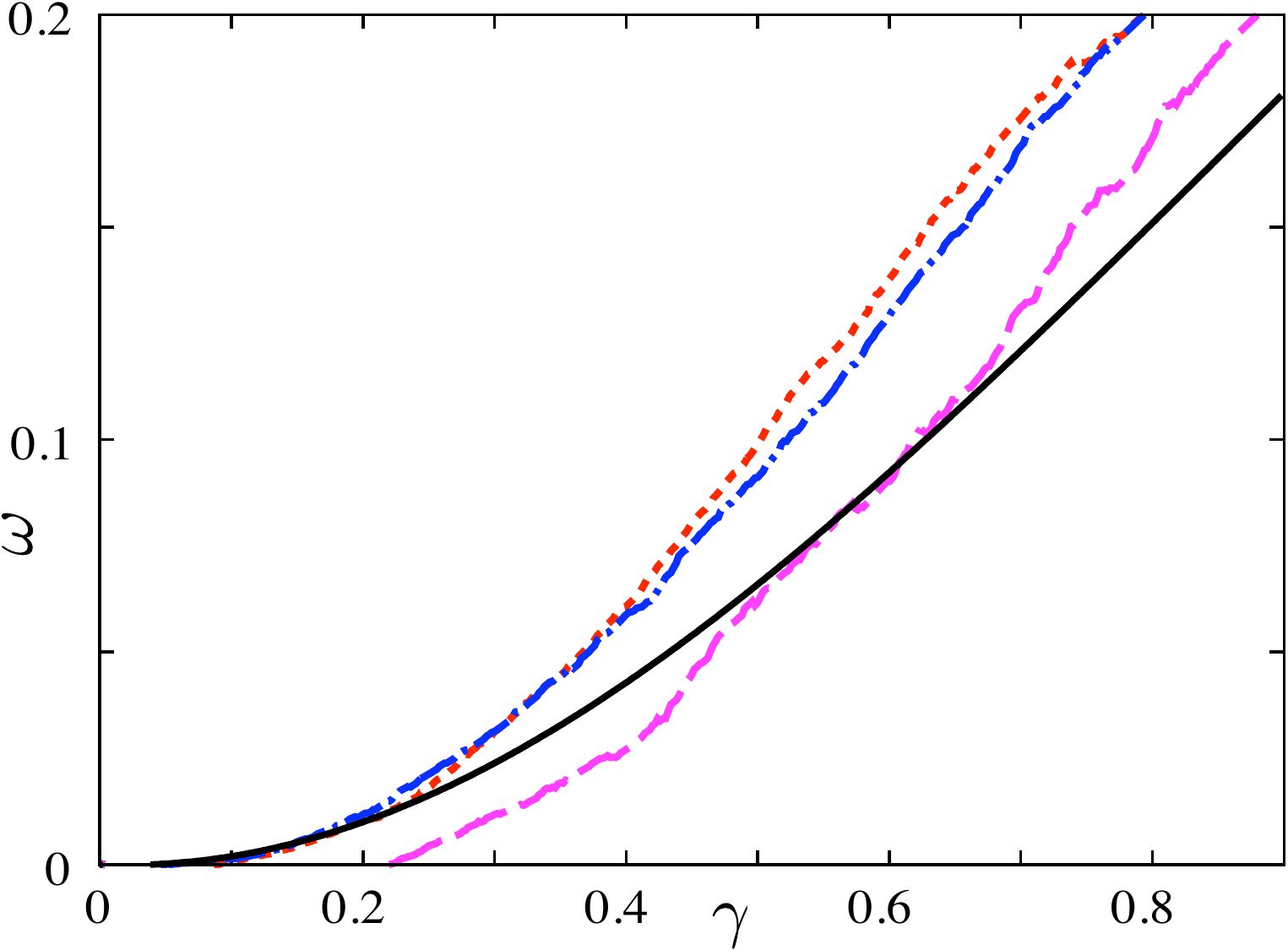}
\includegraphics[width=0.7\linewidth]{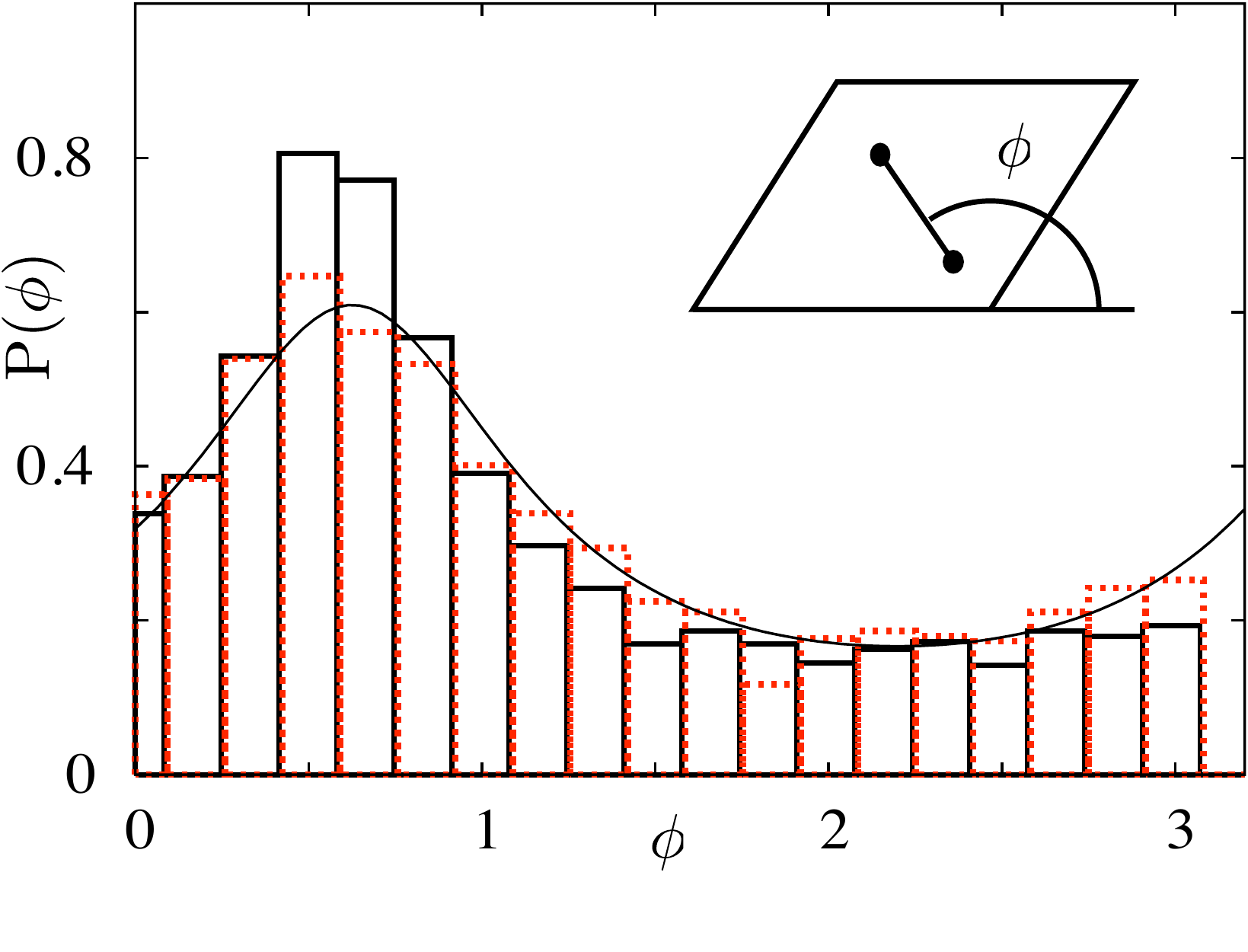}
\end{center}
\caption{(color online) Non-affine behavior of the networks during shearing. {a)} The scaled overall non-affinity ${\sf A}/r_c^2$ at $\gamma=0$ as a function of $L/l_c$ for persistence lengths $l_p/l_c=0.77$ ($\bullet$) and $l_p/l_c=15.7$ ($\ast$). Curves are drawn as a guide to the eye. {b)} Ordering during shearing, after subtracting the value of $\omega$ at $\gamma=0$. The solid line indicates the ordering due to affine shearing. The dotted lines indicate the ordering during shearing in networks with (from bottom to top) $L/l_c=4.0$, $L/l_c=6.0$ and $L/l_c=12.0$. {c)} Distribution of angles with respect to the $x$-axis, both for an affine deformation (dotted bars) and a non-affine deformation (solid bars) at $\gamma=0.7$. The curve is the analytic expression for the angular distribution of an initially isotropic material at shear $\gamma=0.7$. The inset shows the sheared box, in which two connected crosslinks are indicated by dots; their end-to-end vector makes an angle $\phi$ with the $\hat x$-axis.}
\label{fig7b}
\end{figure}

\begin{figure} 
\begin{center}
\includegraphics[width=0.7\linewidth]{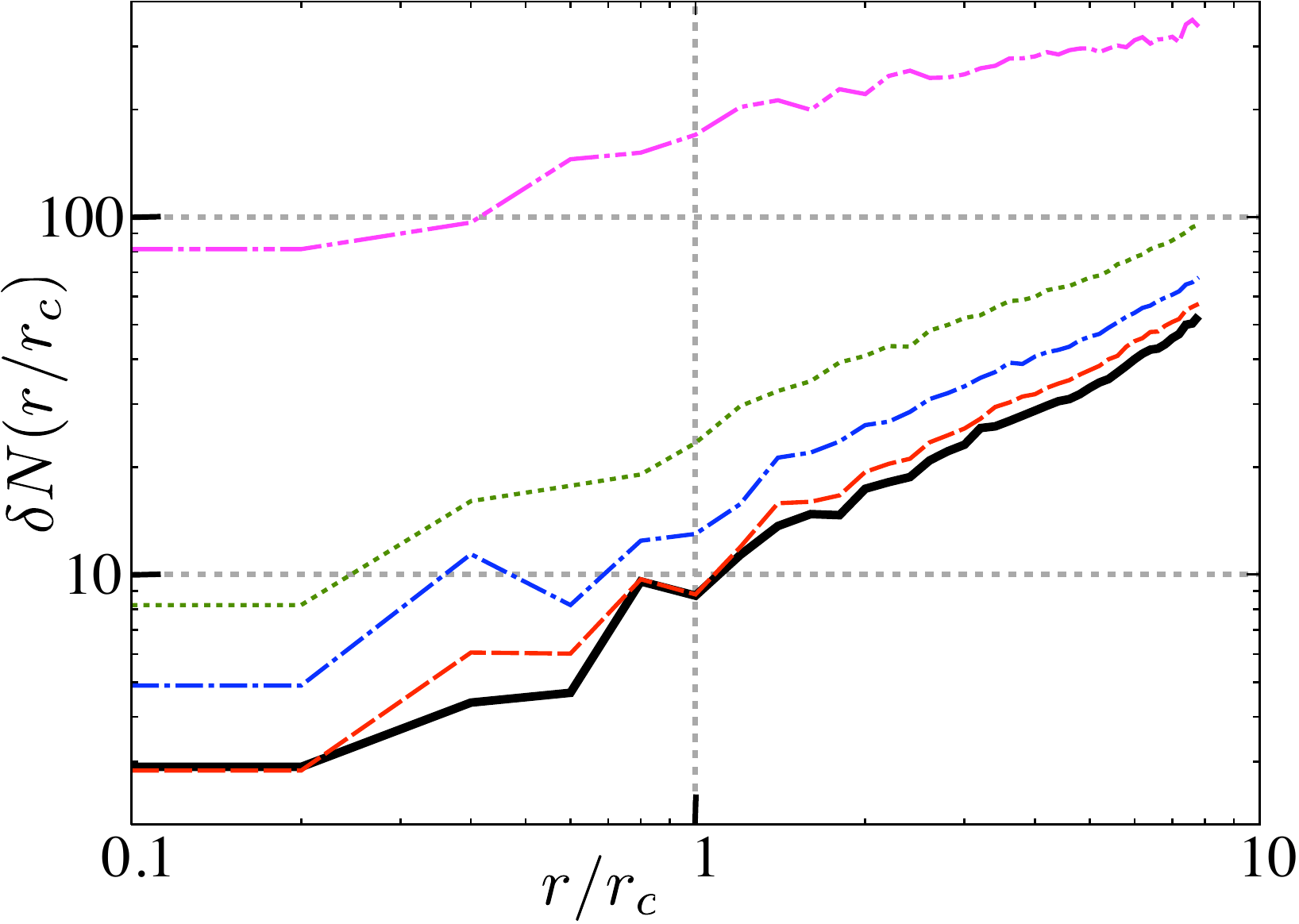}
\includegraphics[width=0.7\linewidth]{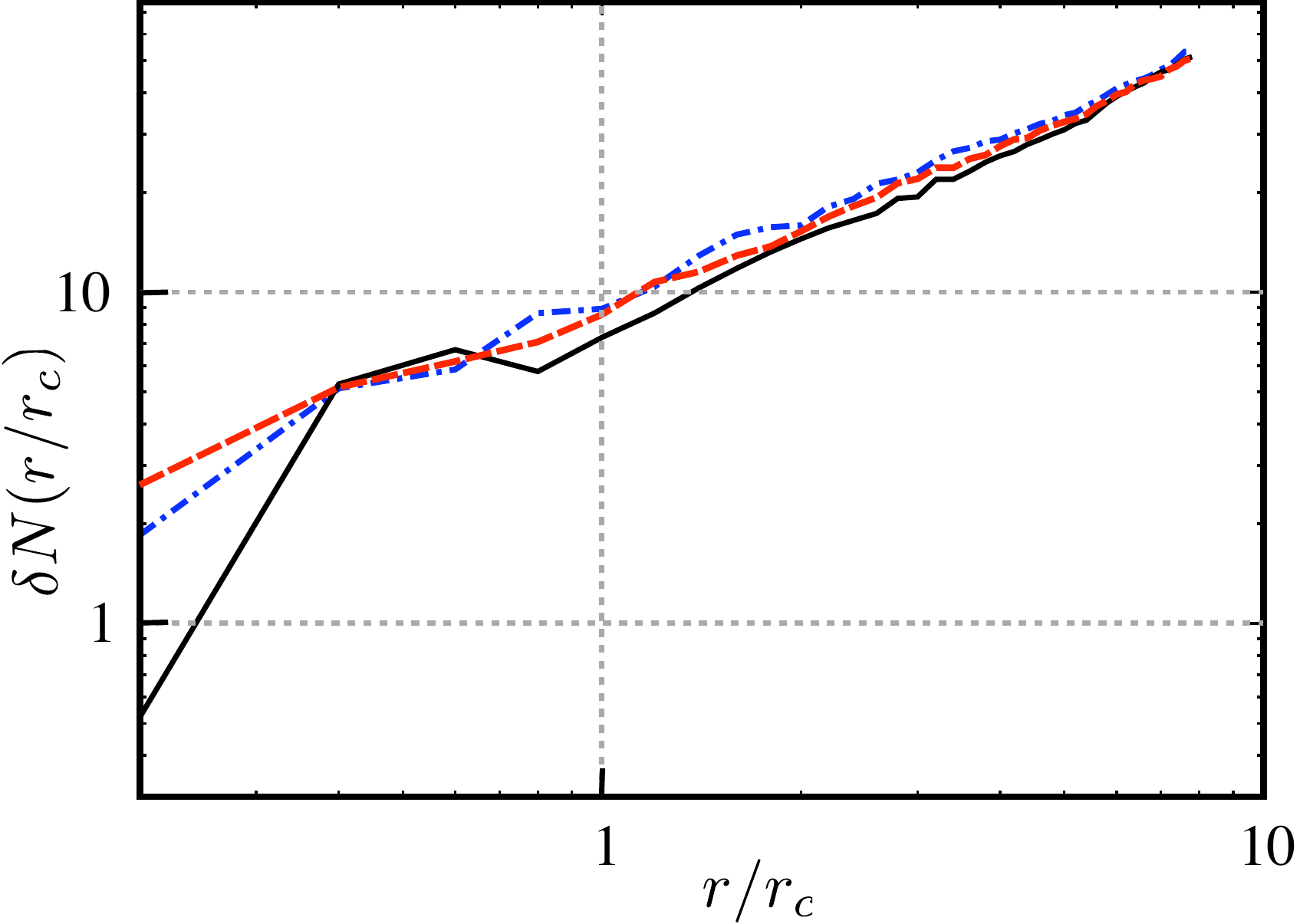}
\includegraphics[width=0.7\linewidth]{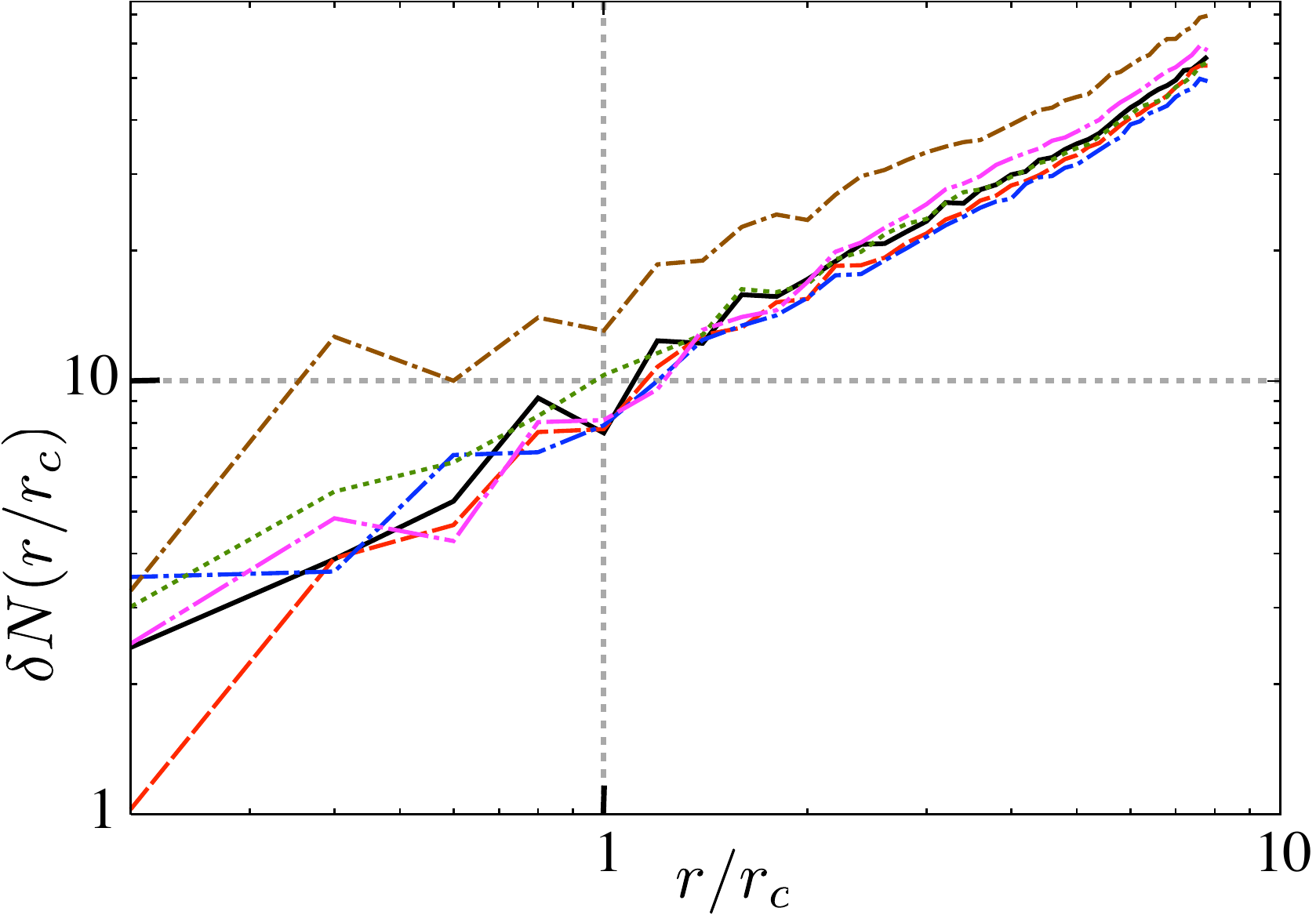}
\end{center}
\caption{(color online) {Correlation $\delta N(r/r_c)$ of non-affine behavior, as a function of distance $r/r_c$}; {a)} for strains $\gamma=0.0$ ($K=K_0$; bottom curve), $\gamma=0.2$ ($K=4.0 K_0$), $\gamma=0.25$ ($K=7.6 K_0$), $\gamma=0.29$ ($K=25 K_0$) and $\gamma=0.33$ ($K=260 K_0$; top curve). {b)} for networks with different $l_p/l_c$ (solid line: $l_p/l_c=0.77$, dash-dotted line: $l_p/l_c=3.81$ and dashed line: $l_p/l_c=15.7$); for all networks, $L/l_c=6$. {c)} for networks with different $L/l_c$, ranging from $4$ to $20$. All curves collapse, except the dash-dotted (brown) curve with $L/l_c=4$. For all networks, $l_p/l_c=15.7$.}
\label{fig8}
\end{figure}
To see whether our systems tend to affinity at the largest strains, we measure the differential non-affinity $\delta {\sf A}$ as a function of the applied macroscopic strain. To relate $\delta {\sf A}$ to other length scales in the system, we plot $\delta {\sf A}/r_c^2$, where $r_c$ is the average distance between crosslinks. Fig.~\ref{fig7a}b shows a strong increase in $\delta {\sf A}$ with increasing strain. From the inset of Fig.~\ref{fig7a}b, a strong correlation between the stiffening of the networks and the amount of non-affine displacements is revealed. Apparently, to prevent the extreme extension that filaments would experience at high strains in an affine setting, the network shows a strong non-affine reorientation. As indicated before, we expect that for high shear all filaments will be aligned in the direction of shear and deform purely by stretching. Since stretching is an affine deformation, we expect $\delta {\sf A}$ to ultimately tend to zero at large strain. This figure and the inset make it clear that this asymptotically (differentially) affine regime is never actually attained, and non-affinity will continue to feature prominently all the way up to the point of failure.

Shorter filaments, or filaments that are less densely crosslinked, are less constrained in their motions which should, in principle, allow for greater non-affine motions. To verify whether indeed this is the case, we evaluate the non-affinity as a function of filament length. To better compare to existing experiments, we shall use ${\sf A}$, the overall non-affinity parameter, instead of the differential measure $\delta {\sf A}$. Fig.~\ref{fig7b}a indeed shows a pronounced increase in non-affinity as the length decreases. This is in agreement with experiments on f-actin, which also show an increase in ${\sf A}(0)$ for decreasing filament length \cite{Koenderink}. Translating ${\sf A}(0)$ to real distances gives for actin with an average $l_c=1 \mu m$ values between $2-6 \mu m^2$, which is close to the values between $2-10 \mu m^2$ reported for experiments \cite{Koenderink}. Besides the length dependence, we also observe a dependency of $l_p$ on the non-affinity: networks of stiffer filaments behave more non-affinely. This suggests that asymptotically, we recover the classical picture of rubber elasticity (which does very well for long, flexible polymers but rather poorly for semiflexible systems): as the persistence length decreases, the polymer configurations become increasingly random (i.e., Gaussian) which is accompanied by a decrease in the non-affinity. This is precisely the rubber limit: Gaussian polymers deforming affinely. Note that these non-affine deformations are the sole possible origin of the large difference in stiffness between affine and non-affine deformations shown above. Thus, even though the magnitude of the non-affine deformations is small, they do have an important effect on the network response. This is a striking example of the value of simulations in this field: microscopic structure and motion are of crucial importance to properly understand the macroscopic behavior.

Apparently, the macroscopic result of the microscopic non-affinity is to lower the overall stiffness of the system. This suggests an interesting question: if the filaments do not go to their affine positions, where {\em do} they go? To begin to answer this, we consider the orientational order of our networks and compute the nematic order parameter $\omega$, defined as $\omega=\langle 3\cos^2\theta-1 \rangle/2$. Here the average is taken over all vectors connecting crosslinks that are connected by segments of filaments and $\theta$ is the angle between such a vector and the average orientation. An isotropic network has $\omega=0$, while a fully ordered network has $\omega=1$. Even when we shear a network affinely, the order will increase from zero to one. To appreciate the effect non-affinity has, we should therefore compare to the affine ordering. This affine ordering is represented by the solid line in Fig.~\ref{fig7b}b. The dotted lines show the effect of non-affine reorientations on the ordering of the network, for different filament lengths. Interestingly, non-affine reorientations tend to increase the order in the network, a behavior independent of $l_p$.

To get insight in the direction of the ordering, we plot the distribution of the angle $\phi$ of the end-to-end vectors of segments with respect to the $x$-axis at $\gamma=0.7$, as shown in Fig.~\ref{fig7b}c. By comparing the distribution in a non-affine network deformation (straight bars) with the distribution of an affine network deformation (dotted bars) we see that the non-affinity increases the number of segments oriented at a small angle. To appreciate the differences in the two distributions, we plot the analytic expression for the distribution of an initial isotropic medium that is sheared affinely. Interestingly, the maximum of $P(\phi)$ coincides with the maximal extensional strain experienced as a function of angle. As the figure clearly shows, the additional ordering is in the direction of maximal extensional strain. This might seem counterintuitive - the order appears to be increasing in the direction of increasing filament extensional strain, which would be highly unfavorable from an energetic point of view. However, one should keep in mind that non-affine motions are not purely rotational: they may encompass additional and simultaneous overall shifts and extensional/compressional components. It would be most interesting to see if this increased order is also observed in experiments. Our simulations suggest that systems containing long filaments are the best place to look for this effect, even though these tend to display lower overall non-affinity.

So far, we have considered only the non-affine motion of single points. The non-affinity correlation function is not only a function of strain, it may also be evaluated for spatially separated points ${\bm x}$ and ${\bm x}'$. To this end, we consider $\delta N(r)=\langle(\vec{r}-\vec{r}_{\rm aff})^2\rangle_r/\delta \gamma^2$, where $\vec{r}$ is the actual vector between two crosslinks and $\vec{r}_{\rm aff}$ is the vector between the crosslinks if they would have moved affine during $\delta\gamma$. The average, now, runs over all pairs of crosslinks whose separation is $r$.

As explained in ref. \cite{Koenderink}, there are two limiting cases in the behavior of $\delta N(r)$. If filaments would be stiff rods, the only way to adapt to strain would be by rotating the whole filament. In that case, doubling $r$ would double $\vec{r}-\vec{r}_{\rm aff}$ and thus $\delta N(r)\sim r^2$. In the other limiting case, segments along a filament behave totally uncorrelated, leading to $\delta N(r)\sim r^0$. The latter is also the limit for $r\rightarrow\infty$. However, the net effect of correlated motion of segments along a filament will be highly sensitive to the actual network configuration.

Fig.~\ref{fig8}a shows $\delta N(r)$ of a network at different strains. Note that the larger scatter for small values of $r$ is due to the smaller number of pairs of crosslinks. As can be seen, $\delta N(r)$ varies with strain: low strains give a low initial value of $\delta N(r)$ and a steep increase while high strains show just the opposite. This behavior is strongly correlated to the stiffening behavior shown in Fig.~\ref{fig5}a (upper line). The observed strain-dependence of $\delta N(r)$ is indiscernible when normalizing with respect to $\gamma$ rather than of $\delta \gamma$, which might explain why experiments report no strain-dependence \cite{Koenderink}.

We observe a small but systematic dependence on $l_p$, as shown in Fig.~\ref{fig8}b. Interestingly, thus far we hardly observe any length dependence of the correlation in non-affine behavior, which is shown in Fig.~\ref{fig8}c for networks with $l_p/l_c=15.7$. For filaments with $l_p>l_c$, one would naively expect that the behavior of segments along a filament will be much more correlated than the behavior of segments belonging to different filaments. Thus, one might expect to find increasing spatial correlations for systems composed of larger filaments. That we do not see this behavior suggests that it is approximately balanced by another effect: larger filaments have more links to the rest of the system and are therefore more constrained. While the individual segments along a single filament would like to line up, they become increasingly unable to do so. Interestingly, the first experiments to measure $\delta N(r)$ do show a length dependence \cite{Koenderink}. We cannot rule out that we will see this behavior at larger system sizes, but for now are unable to reproduce it.

\begin{figure}[ht] 
\begin{center}
\includegraphics[width=0.6\linewidth]{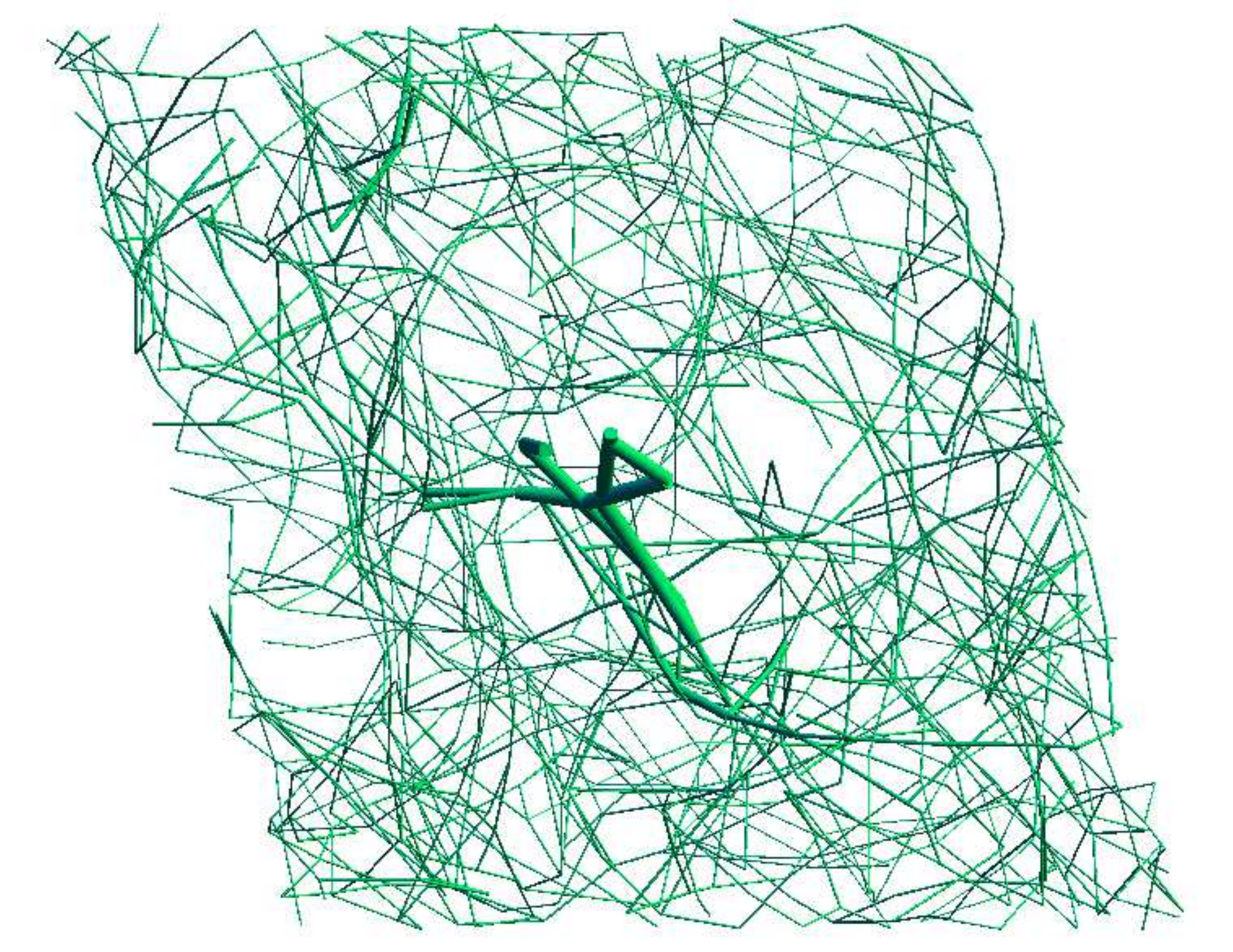}
\includegraphics[width=0.6\linewidth]{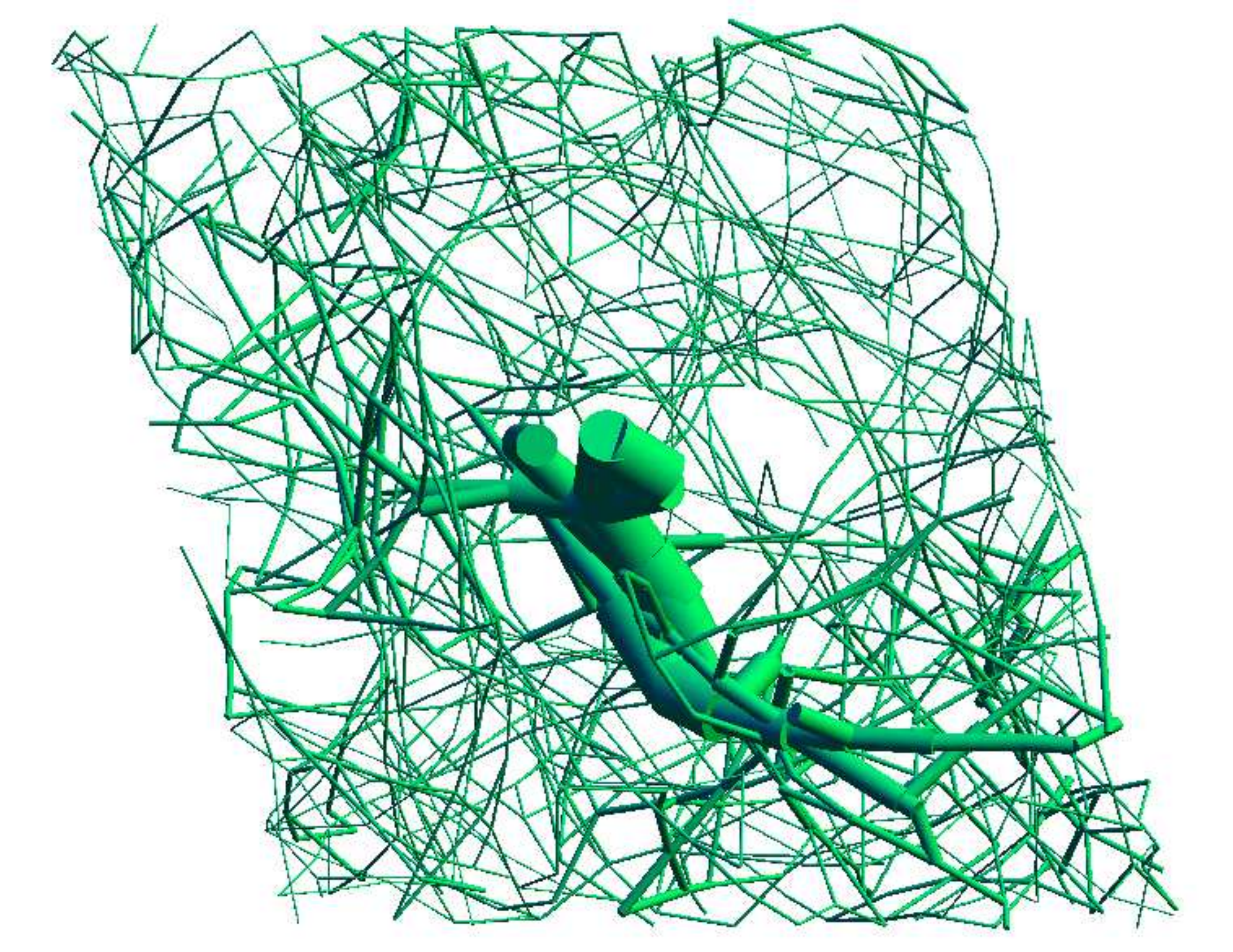}
\includegraphics[width=0.6\linewidth]{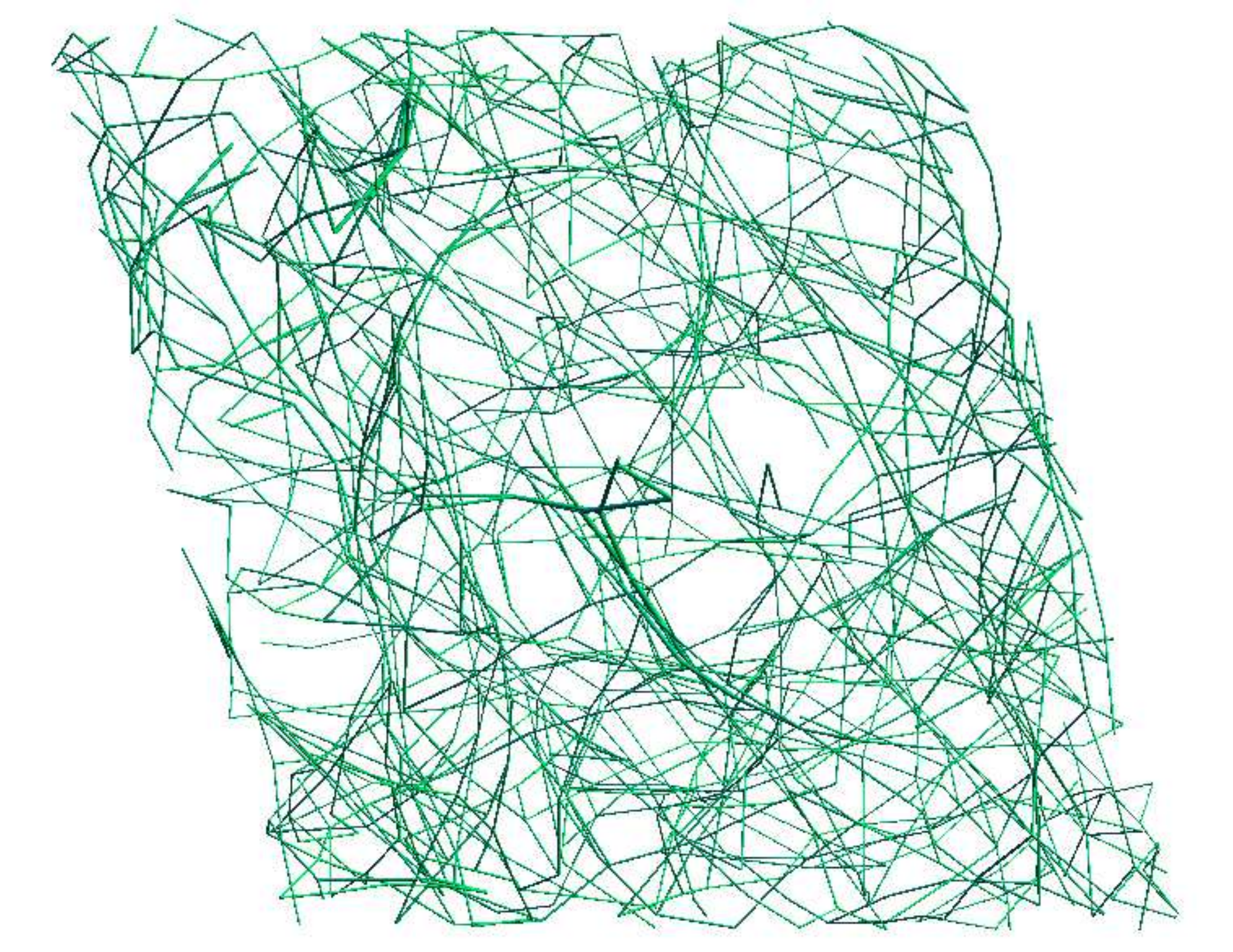}
\end{center}
\caption{Illustration of a collective reorientation in one of our networks with $l_p/l_c=15.7$ and $L/l_c=6$ during deformation at respectively $\gamma = 0.268$, $\gamma = 0.270$ and $\gamma = 0.272$. The thickness of the segment indicates the size of its displacement. The modulus $K$ belonging to the deformation of this network is shown as the upper curve in the inset in Fig.~\ref{fig5}a.}
\label{fig9}
\end{figure}

\subsection{Collective Rearrangements}

A closer look at Fig.~\ref{fig5}a reveals some outliers in the $K$ vs. $\gamma$ curve. These discontinuities in $K$ are accompanied by an increase in $A$. This is not a glitch, but rather reflects an interesting microscopic aspect of our networks. To understand this behavior we look at the displacements of individual segments during shearing. Fig.~\ref{fig9} shows a network in which the thickness of the segments indicates their displacement during a strain increment of $0.2\%$. Here we see what happens: during such a strain increment, a significant fraction of the segments has a relatively large incremental displacement in comparison with the average displacement of segments during shear increments.

The noteworthy feature is not so much that there are large displacements, but rather that these displacements are localized and occur in correlated fashion. This is reminiscent of the behavior of so-called collectively rearranging regions, observed in simulation and experiment in glassy systems and colloidal suspensions. These events are rare over the time courses that we have simulated, but may turn out to play an important role in the long-time behavior of these materials. It would be most interesting to check whether these events are also seen in experiments - while these may not be able to resolve the blip in $K$ they might be able to register the accompanying peaks in $A$. The weight in determining $A$ of a reorientation of a certain size decreases with increasing $\gamma$, since $A$ measures the total non-affinity relative to the total shear. This implies that for small shears, reorientations might induce huge peaks in $A$, while these peaks are absent for larger shears even though the reorientations are still present.

\section{Conclusions}
We have presented a new method to generate and deform 3D networks of biopolymer filaments. By an adequate choice of energies both the entropic stiffness of individual segments as well as the persistence of filaments through crosslinks can be taken into account. By a Monte Carlo thermalization the networks find a local minimum, without further interference from our side.

This method enables us to relate the macroscopic network response to microscopic behavior of individual segments and crosslinks, both at small and large strains. Although a quantitative comparison between experiments and our simulations is hard to obtain, the first results from these simulations agree well with experiments. Both the stiffening during shearing and the length-dependency of the non-affinity are as expected and fit well into the general framework of the behavior of semiflexible polymers. Besides, the stress-dependence of the stiffness for large shears is the same as experiments have shown. This confirms that our model captures the right features that decide the network behavior.

Our model proves an excellent tool to compare affine deformations with deformations that allow for non-affine displacements. We have shown that non-affine displacements have a large influence on the stiffness of a network and the onset of stiffening. This accounts for the important role of filament length. Besides, the accuracy of analysis of the behavior of the filaments during deformation reveals some surprising results that are hard to obtain by experimental analysis. Thus far unobserved, the non-affinity increases the order in the networks.

Thus far we have only considered networks of a single type of filaments. Both in cells and in the extracellular matrix, the important load-bearing biopolymer networks are made up of different kinds of filaments: vessel walls are composed of collagen and elastin, and the cytoskeleton too is a composite system containing f-actin, intermediate filaments and microtubules. This method is a promising tool to explore the behavior of such composite networks under strain, and we are currently exploring their properties.

\section{Acknowledgements}
It is a pleasure to acknowledge Fred MacKintosh for many helpful discussions.

\end{document}